\documentclass[11pt]{article}
\usepackage{axodraw}
\newcommand{\la}[1]{\label{#1}}
\newcommand{\be}{\begin{equation}}
\newcommand{\ee}{\end{equation}}
\newcommand{\ba}{\begin{eqnarray}}
\newcommand{\ea}{\end{eqnarray}}
\newcommand{\rmi}[1]{{\mbox{\scriptsize #1}}}

\newcommand{\eq}{Eq.~}
\newcommand{\se}{Sec.~}
\newcommand{\eqs}{Eqs.~}
\newcommand{\nr}[1]{(\ref{#1})}

\newcommand{\nn}{\nonumber \\}
\newcommand{\fr}[2]{{\frac{#1}{#2}\,}}

\renewcommand{\(}{\left(}
\renewcommand{\)}{\right)}
\newcommand{\lb}{\left\{}
\newcommand{\rb}{\right\}}
\newcommand{\lk}{\left[}
\newcommand{\rk}{\right]}
\newcommand{\ld}{\left.}

\renewcommand{\d}{\delta}

\newcommand{\6}{\partial}
\newcommand{\Av}[1]{\left\langle #1 \right\rangle}

\newcommand{\sy}[3]{{\textstyle #1\frac{#2}{#3}}}

\def\plus{+} 
\newcommand{\sm}[1]{\,{\scriptstyle #1}\,}

\makeatletter
\renewcommand\section{\@startsection {section}{1}{\z@}%
                                   {-5.5ex \@plus -1ex \@minus -.2ex}
                                   {2.3ex \@plus.2ex}%
                                   {\normalfont\large\bfseries}}
\renewcommand\subsection{\@startsection{subsection}{2}{\z@}%
                                     {-3.25ex\@plus -1ex \@minus -.2ex}%
                                     {1.5ex \@plus .2ex}%
                                     {\normalfont\normalsize\bfseries}}
\renewcommand\thesection {\@arabic\c@section}
\renewcommand\thesubsection   {\thesection.\@arabic\c@subsection}
\renewcommand{\@seccntformat}[1]{%
\csname the#1\endcsname.\hspace{1.0em}}
\makeatother

\newcommand{\pic}[1]{\;\parbox[c]{30pt}{\begin{picture}(30,30)(0,0)
\SetWidth{1.0}\SetScale{1.0} #1 \end{picture}}\;}
\newcommand{\pib}[1]{\;\parbox[c]{36pt}{\begin{picture}(22.5,30)(0,0)
\SetWidth{1.0}\SetScale{1.0} #1 \end{picture}}\;}
\newcommand{\picb}[1]{\;\parbox[c]{45pt}{\begin{picture}(45,30)(0,0)
\SetWidth{1.0}\SetScale{1.0} #1 \end{picture}}\;}
\newcommand{\picc}[1]{\;\parbox[c]{60pt}{\begin{picture}(60,30)(0,0)
\SetWidth{1.0}\SetScale{1.0} #1 \end{picture}}\;}

\newcommand{\spic}[1]{\;\parbox[c]{21pt}{\begin{picture}(21,21)(0,0)
\SetWidth{1.0}\SetScale{0.7} #1 \end{picture}}\;}

\newcommand{\spicb}[1]{\;\parbox[c]{32pt}{\begin{picture}(32,21)(0,0)
\SetWidth{1.0}\SetScale{0.7} #1 \end{picture}}\;}
\newcommand{\spicc}[1]{\;\parbox[c]{42pt}{\begin{picture}(42,21)(0,0)
\SetWidth{1.0}\SetScale{0.7} #1 \end{picture}}\;}

\newcommand{\spiccc}[1]{\;\parbox[c]{63pt}{\begin{picture}(63,21)(0,0)
\SetWidth{1.0}\SetScale{0.7} #1 \end{picture}}\;}

\def\Lwidth{1}

\def\Agl(#1,#2)(#3,#4,#5){\PhotonArc(#1,#2)(#3,#4,#5){\Lwidth}
{6.283 #3 mul 360 div #4 #5 sub #4 #5 sub mul sqrt mul Ldensity mul}}
\def\Lgl(#1,#2)(#3,#4){\Photon(#1,#2)(#3,#4){\Lwidth}
{#1 #3 sub #1 #3 sub mul #2 #4 sub #2 #4 sub mul add sqrt Ldensity mul}}
\def\Agh(#1,#2)(#3,#4,#5){\DashArrowArc(#1,#2)(#3,#4,#5){1}}
\def\Aagh(#1,#2)(#3,#4,#5){\DashArrowArcn(#1,#2)(#3,#5,#4){1}}
\def\Lgh(#1,#2)(#3,#4){\DashArrowLine(#1,#2)(#3,#4){1}}
\def\Lagh(#1,#2)(#3,#4){\DashArrowLine(#3,#4)(#1,#2){1}}
\def\Ahh(#1,#2)(#3,#4,#5){\DashCArc(#1,#2)(#3,#4,#5){1}}
\def\Lhh(#1,#2)(#3,#4){\DashLine(#1,#2)(#3,#4){1}}
\def\Aqu(#1,#2)(#3,#4,#5){\ArrowArc(#1,#2)(#3,#4,#5)}
\def\Aaqu(#1,#2)(#3,#4,#5){\ArrowArcn(#1,#2)(#3,#5,#4)}
\def\Lqu(#1,#2)(#3,#4){\ArrowLine(#1,#2)(#3,#4)}
\def\Laqu(#1,#2)(#3,#4){\ArrowLine(#3,#4)(#1,#2)}
\def\Aqq(#1,#2)(#3,#4,#5){\CArc(#1,#2)(#3,#4,#5)}
\def\Lqq(#1,#2)(#3,#4){\Line(#1,#2)(#3,#4)}
\def\Asc(#1,#2)(#3,#4,#5){\CArc(#1,#2)(#3,#4,#5)}
\def\Lsc(#1,#2)(#3,#4){\Line(#1,#2)(#3,#4)}
\def\DAsc(#1,#2)(#3,#4,#5){\DashCArc(#1,#2)(#3,#4,#5){3}}
\def\DLsc(#1,#2)(#3,#4){\DashLine(#1,#2)(#3,#4){3}}
\def\TAsc(#1,#2)(#3,#4,#5){\SetWidth{2.0}\CArc(#1,#2)(#3,#4,#5)\SetWidth{1.0}}
\def\TLsc(#1,#2)(#3,#4){\SetWidth{2.0}\Line(#1,#2)(#3,#4)\SetWidth{1.0}}

\def\TopoVR(#1){\pic{#1(15,15)(15,-90,270)}}
\def\TopoVRoo(#1){\;\pic{#1(15,15)(15,0,180) #1(15,15)(15,180,360)%
 \GCirc(0,15){5}{0.75} \GCirc(30,15){5}{0.75}}\;}
\def\TopoVRor(#1){\;\pic{#1(15,15)(15,0,180) #1(15,15)(15,180,360)%
 \GCirc(0,15){5}{0.75} \BBoxc(30,15)(6,6)}\;}
\def\TopoVRoi(#1){\;\pic{#1(15,15)(15,0,180) #1(15,15)(15,180,360)%
 \GCirc(0,15){5}{0.75} \GBoxc(30,15)(6,6){0}}\;}
\def\TopoVRooo(#1){\pic{#1(15,15)(15,-30,90) #1(15,15)(15,90,210)%
 #1(15,15)(15,210,330) \GCirc(15,30){3}{0} \GCirc(2,7.5){3}{0}%
 \GCirc(28,7.5){3}{0}}}
\def\TTopoVRoo(#1){\;\pic{#1(15,15)(15,0,180) #1(15,15)(15,180,360)%
 \GCirc(0,15){5}{0.75} \Text(0,15)[c]{$\scriptstyle 1$}
 \GCirc(30,15){5}{0.75} \Text(30,15)[c]{$\scriptstyle 1$}}\;}
\def\TTopoVRor(#1){\;\pic{#1(15,15)(15,0,180) #1(15,15)(15,180,360)%
 \GCirc(0,15){5}{0.75} \Text(0,15)[c]{$\scriptstyle 1$}
 \GBoxc(30,15)(9,9){0.75} \Text(30,15)[c]{$\scriptstyle 2$} }\;}
\def\TTopoVRoi(#1){\;\pic{#1(15,15)(15,0,180) #1(15,15)(15,180,360)%
 \GCirc(0,15){5}{0.75} \Text(0,15)[c]{$\scriptstyle 1$} 
 \GCirc(30,15){5}{0.75} \Text(30,15)[c]{$\scriptstyle 2$} }\;}
\def\TTopoVRooo(#1){\pic{#1(15,15)(15,-30,90) #1(15,15)(15,90,210)%
 #1(15,15)(15,210,330) 
 \GCirc(15,30){5}{0.75}\Text(15,30)[c]{$\scriptstyle 1$}%
 \GCirc(2,7.5){5}{0.75}\Text(2,7.5)[c]{$\scriptstyle 1$}%
 \GCirc(28,7.5){5}{0.75}\Text(28,7.5)[c]{$\scriptstyle 1$}}}
\def\TTopoVRoooo(#1){\pic{#1(15,15)(15,-45,45) #1(15,15)(15,45,135)%
 #1(15,15)(15,135,225) #1(15,15)(15,225,315) 
 \GCirc(25,25){5}{0.75}\Text(25,25)[c]{$\scriptstyle 1$}%
 \GCirc(25,5){5}{0.75}\Text(25,5)[c]{$\scriptstyle 1$}%
 \GCirc(5,25){5}{0.75}\Text(5,25)[c]{$\scriptstyle 1$}%
 \GCirc(5,5){5}{0.75}\Text(5,5)[c]{$\scriptstyle 1$}}}
\def\TTopoVRooi(#1){\pic{#1(15,15)(15,-30,90) #1(15,15)(15,90,210)%
 #1(15,15)(15,210,330) 
 \GCirc(15,30){5}{0.75}\Text(15,30)[c]{$\scriptstyle 2$}%
 \GCirc(2,7.5){5}{0.75}\Text(2,7.5)[c]{$\scriptstyle 1$}%
 \GCirc(28,7.5){5}{0.75}\Text(28,7.5)[c]{$\scriptstyle 1$}}}
\def\TTopoVRoor(#1){\pic{#1(15,15)(15,-30,90) #1(15,15)(15,90,210)%
 #1(15,15)(15,210,330) 
 \GBoxc(15,30)(9,9){0.75}\Text(15,30)[c]{$\scriptstyle 2$}%
 \GCirc(2,7.5){5}{0.75}\Text(2,7.5)[c]{$\scriptstyle 1$}%
 \GCirc(28,7.5){5}{0.75}\Text(28,7.5)[c]{$\scriptstyle 1$}}}
\def\TTopoVRiie(#1){\;\pic{#1(15,15)(15,0,180) #1(15,15)(15,180,360)%
 \GCirc(0,15){5}{0.75} \Text(0,15)[c]{$\scriptstyle 2$} 
 \GCirc(30,15){5}{0.75} \Text(30,15)[c]{$\scriptstyle 2$} }\;}
\def\TTopoVRire(#1){\;\pic{#1(15,15)(15,0,180) #1(15,15)(15,180,360)%
 \GCirc(0,15){5}{0.75} \Text(0,15)[c]{$\scriptstyle 2$}
 \GBoxc(30,15)(9,9){0.75} \Text(30,15)[c]{$\scriptstyle 2$} }\;}
\def\TTopoVRiif(#1){\;\pic{#1(15,15)(15,0,180) #1(15,15)(15,180,360)%
 \GCirc(0,15){5}{0.75} \Text(0,15)[c]{$\scriptstyle 1$} 
 \GCirc(30,15){5}{0.75} \Text(30,15)[c]{$\scriptstyle 3$} }\;}
\def\TTopoVRirf(#1){\;\pic{#1(15,15)(15,0,180) #1(15,15)(15,180,360)%
 \GCirc(0,15){5}{0.75} \Text(0,15)[c]{$\scriptstyle 1$}
 \GBoxc(30,15)(9,9){0.75} \Text(30,15)[c]{$\scriptstyle 3$} }\;}
\def\TTopoVRirrf(#1){\;\pic{#1(15,15)(15,0,180) #1(15,15)(15,180,360)%
 \GCirc(0,15){5}{0.75} \Text(0,15)[c]{$\scriptstyle 1$}
 \GBoxc(30,15)(12,12){0.75}
 \GBoxc(30,15)(8,8){0.75} \Text(30,15)[c]{$\scriptstyle 3$} }\;}
\def\STTopoVRoo(#1){\;\spic{#1(15,15)(15,0,180) #1(15,15)(15,180,360)%
 \GCirc(0,15){5}{0.75} \Text(0,10.5)[c]{$\scriptscriptstyle 1$}
 \GCirc(30,15){5}{0.75} \Text(21,10.5)[c]{$\scriptscriptstyle 1$}}\;}
\def\STTopoVRor(#1){\;\spic{#1(15,15)(15,0,180) #1(15,15)(15,180,360)%
 \GCirc(0,15){5}{0.75} \Text(0,10.5)[c]{$\scriptscriptstyle 1$}
 \GBoxc(30,15)(9,9){0.75} \Text(21,10.5)[c]{$\scriptscriptstyle 2$} }\;}
\def\STTopoVRoi(#1){\;\spic{#1(15,15)(15,0,180) #1(15,15)(15,180,360)%
 \GCirc(0,15){5}{0.75} \Text(0,10.5)[c]{$\scriptscriptstyle 1$} 
 \GCirc(30,15){5}{0.75} \Text(21,10.5)[c]{$\scriptscriptstyle 2$} }\;}
\def\STTopoVRooo(#1){\spic{#1(15,15)(15,-30,90) #1(15,15)(15,90,210)%
 #1(15,15)(15,210,330) 
 \GCirc(15,30){5}{0.75}\Text(10.5,21)[c]{$\scriptscriptstyle 1$}%
 \GCirc(2,7.5){5}{0.75}\Text(1.4,5.25)[c]{$\scriptscriptstyle 1$}%
 \GCirc(28,7.5){5}{0.75}\Text(19.6,5.25)[c]{$\scriptscriptstyle 1$}}}
\def\STTopoVRoooo(#1){\spic{#1(15,15)(15,-45,45) #1(15,15)(15,45,135)%
 #1(15,15)(15,135,225) #1(15,15)(15,225,315) 
 \GCirc(25,25){5}{0.75}\Text(17.5,17.5)[c]{$\scriptscriptstyle 1$}%
 \GCirc(25,5){5}{0.75}\Text(17.5,3.5)[c]{$\scriptscriptstyle 1$}%
 \GCirc(5,25){5}{0.75}\Text(3.5,17.5)[c]{$\scriptscriptstyle 1$}%
 \GCirc(5,5){5}{0.75}\Text(3.5,3.5)[c]{$\scriptscriptstyle 1$}}}
\def\STTopoVRooi(#1){\spic{#1(15,15)(15,-30,90) #1(15,15)(15,90,210)%
 #1(15,15)(15,210,330) 
 \GCirc(15,30){5}{0.75}\Text(10.5,21)[c]{$\scriptscriptstyle 2$}%
 \GCirc(2,7.5){5}{0.75}\Text(1.4,5.25)[c]{$\scriptscriptstyle 1$}%
 \GCirc(28,7.5){5}{0.75}\Text(19.6,5.25)[c]{$\scriptscriptstyle 1$}}}
\def\STTopoVRoor(#1){\spic{#1(15,15)(15,-30,90) #1(15,15)(15,90,210)%
 #1(15,15)(15,210,330) 
 \GBoxc(15,30)(9,9){0.75}\Text(10.5,21)[c]{$\scriptscriptstyle 2$}%
 \GCirc(2,7.5){5}{0.75}\Text(1.4,5.25)[c]{$\scriptscriptstyle 1$}%
 \GCirc(28,7.5){5}{0.75}\Text(19.6,5.25)[c]{$\scriptscriptstyle 1$}}}
\def\STTopoVRiie(#1){\;\spic{#1(15,15)(15,0,180) #1(15,15)(15,180,360)%
 \GCirc(0,15){5}{0.75} \Text(0,10.5)[c]{$\scriptscriptstyle 2$} 
 \GCirc(30,15){5}{0.75} \Text(21,10.5)[c]{$\scriptscriptstyle 2$} }\;}
\def\STTopoVRire(#1){\;\spic{#1(15,15)(15,0,180) #1(15,15)(15,180,360)%
 \GCirc(0,15){5}{0.75} \Text(0,10.5)[c]{$\scriptscriptstyle 2$}
 \GBoxc(30,15)(9,9){0.75} \Text(21,10.5)[c]{$\scriptscriptstyle 2$} }\;}
\def\STTopoVRiif(#1){\;\spic{#1(15,15)(15,0,180) #1(15,15)(15,180,360)%
 \GCirc(0,15){5}{0.75} \Text(0,10.5)[c]{$\scriptscriptstyle 1$} 
 \GCirc(30,15){5}{0.75} \Text(21,10.5)[c]{$\scriptscriptstyle 3$} }\;}
\def\STTopoVRirf(#1){\;\spic{#1(15,15)(15,0,180) #1(15,15)(15,180,360)%
 \GCirc(0,15){5}{0.75} \Text(0,10.5)[c]{$\scriptscriptstyle 1$}
 \GBoxc(30,15)(9,9){0.75} \Text(21,10.5)[c]{$\scriptscriptstyle 3$} }\;}
\def\STTopoVRirrf(#1){\;\spic{#1(15,15)(15,0,180) #1(15,15)(15,180,360)%
 \GCirc(0,15){5}{0.75} \Text(0,10.5)[c]{$\scriptscriptstyle 1$}
 \GBoxc(30,15)(12,12){0.75}
 \GBoxc(30,15)(8,8){0.75} \Text(21,10.5)[c]{$\scriptscriptstyle 3$} }\;}
\def\TopoVRooT(#1,#2,#3){\;\pic{#1(15,15)(15,0,180) #1(15,15)(15,180,360)%
 \GCirc(0,15){3}{0} \GCirc(30,15){3}{0} \Text(5,15)[l]{{$\scriptstyle #2$}}%
 \Text(25,15)[r]{{$\scriptstyle #3$}}}\;}
\def\TopoVRorT(#1,#2,#3){\;\pic{#1(15,15)(15,0,180) #1(15,15)(15,180,360)%
 \GCirc(0,15){3}{0} \BBoxc(30,15)(6,6) \Text(5,15)[l]{{$\scriptstyle #2$}}%
 \Text(25,15)[r]{{$\scriptstyle #3$}}}\;}
\def\TopoVRoiT(#1,#2,#3){\;\pic{#1(15,15)(15,0,180) #1(15,15)(15,180,360)%
 \GCirc(0,15){3}{0} \GBoxc(30,15)(6,6){0} \Text(5,15)[l]{{$\scriptstyle #2$}}%
 \Text(25,15)[r]{{$\scriptstyle #3$}}}\;}
\def\TopoVRoooT(#1,#2,#3,#4){\pic{#1(15,15)(15,-30,90) #1(15,15)(15,90,210)%
 #1(15,15)(15,210,330) \GCirc(15,30){3}{0} \GCirc(2,7.5){3}{0}%
 \GCirc(28,7.5){3}{0} \Text(15,25)[t]{{$\scriptstyle #2$}}%
 \Text(6,9)[bl]{{$\scriptstyle #3$}} \Text(24,9)[br]{{$\scriptstyle #4$}}}}
\def\ToptVS(#1,#2,#3){\pic{#1(15,15)(15,0,180) #2(15,15)(15,180,360)%
 #3(30,15)(0,15)}}
\def\SToptVS(#1,#2,#3){\spic{#1(15,15)(15,0,180) #2(15,15)(15,180,360)%
 #3(30,15)(0,15)}}
\def\ToptVSTxt(#1,#2,#3,#4,#5){\;\pic{#1(15,15)(15,0,180)%
 #2(15,15)(15,180,360) #3(30,15)(0,15) \GCirc(0,15){3}{0} \GCirc(30,15){3}{0}%
 \Text(4,17)[bl]{{$\scriptstyle #4$}} \Text(26,13)[tr]{{$\scriptstyle #5$}}}\;}
\def\ToptVE(#1,#2){\picc{#1(15,15)(15,0,360) #2(45,15)(15,-180,180)}}
\def\SToptVE(#1,#2){\spicc{#1(15,15)(15,0,360) #2(45,15)(15,-180,180)}}
\def\ToprVM(#1,#2,#3,#4,#5,#6){\pic{#3(15,15)(15,-30,90) #1(15,15)(15,90,210)%
 #2(15,15)(15,210,330) #5(2,7.5)(15,15) #6(15,15)(15,30) #4(28,7.5)(15,15)}}
\def\SToprVM(#1,#2,#3,#4,#5,#6){\spic{#3(15,15)(15,-30,90)%
 #1(15,15)(15,90,210)%
 #2(15,15)(15,210,330) #5(2,7.5)(15,15) #6(15,15)(15,30) #4(28,7.5)(15,15)}}
\def\ToprVV(#1,#2,#3,#4,#5){\!\!\picb{#2(26.25,15)(15,256,76)%
 #3(30,30)(15,30) #1(18.75,15)(15,104,284) #4(15,30)(22.5,0)%
 #5(30,30)(22.5,0)}\!\!}
\def\SToprVV(#1,#2,#3,#4,#5){\!\!\spicb{#2(26.25,15)(15,256,76)%
 #3(30,30)(15,30) #1(18.75,15)(15,104,284) #4(15,30)(22.5,0)%
 #5(30,30)(22.5,0)}\!\!}
\def\ToprVB(#1,#2,#3,#4){\picb{#1(30,15)(15,-120,120) #2(30,15)(15,120,240)%
 #3(15,15)(15,60,300) #4(15,15)(15,-60,60)}}
\def\SToprVB(#1,#2,#3,#4){\spicb{#1(30,15)(15,-120,120) #2(30,15)(15,120,240)%
 #3(15,15)(15,60,300) #4(15,15)(15,-60,60)}}
\def\TopfVX(#1,#2,#3,#4,#5,#6,#7,#8,#9){\picb{#1(15,15)(15,90,270)%
 #2(30,15)(15,-90,90) #4(30,30)(15,30) #3(15,0)(30,0) #6(15,0)(15,15)%
 #5(15,15)(30,30) #8(15,30)(20,25) #8(25,20)(30,15) #7(30,15)(30,0)%
 #9(15,15)(30,15)}}
\def\STopfVX(#1,#2,#3,#4,#5,#6,#7,#8,#9){\spicb{#1(15,15)(15,90,270)%
 #2(30,15)(15,-90,90) #4(30,30)(15,30) #3(15,0)(30,0) #6(15,0)(15,15)%
 #5(15,15)(30,30) #8(15,30)(20,25) #8(25,20)(30,15) #7(30,15)(30,0)%
 #9(15,15)(30,15)}}
\def\TopfVH(#1,#2,#3,#4,#5,#6,#7,#8,#9){\picb{#1(15,15)(15,90,270)%
 #2(30,15)(15,-90,90) #4(30,30)(15,30) #3(15,0)(30,0) #6(15,0)(15,15)%
 #5(15,15)(15,30) #8(30,30)(30,15) #7(30,15)(30,0) #9(15,15)(30,15)}}
\def\STopfVH(#1,#2,#3,#4,#5,#6,#7,#8,#9){\spicb{#1(15,15)(15,90,270)%
 #2(30,15)(15,-90,90) #4(30,30)(15,30) #3(15,0)(30,0) #6(15,0)(15,15)%
 #5(15,15)(15,30) #8(30,30)(30,15) #7(30,15)(30,0) #9(15,15)(30,15)}}
\def\TopfVW(#1,#2,#3,#4,#5,#6,#7,#8){\pic{#1(15,15)(15,90,180)%
 #3(15,15)(15,180,270) #2(15,15)(15,270,360) #4(15,15)(15,0,90)%
 #5(15,15)(15,30) #7(15,15)(15,0) #6(0,15)(15,15) #8(30,15)(15,15)}}
\def\STopfVW(#1,#2,#3,#4,#5,#6,#7,#8){\spic{#1(15,15)(15,90,180)%
 #3(15,15)(15,180,270) #2(15,15)(15,270,360) #4(15,15)(15,0,90)%
 #5(15,15)(15,30) #7(15,15)(15,0) #6(0,15)(15,15) #8(30,15)(15,15)}}
\def\TopfVV(#1,#2,#3,#4,#5,#6,#7,#8){\!\!\picb{#2(26.25,15)(15,256,346)%
 #3(26.25,15)(15,-14,76) #4(30,30)(15,30) #1(18.75,15)(15,104,284)%
 #7(22.5,0)(15,30) #6(30,30)(26.25,15) #8(26.25,15)(22.5,0)%
 #5(25.25,15)(39.8,11.4)}\!\!}
\def\STopfVV(#1,#2,#3,#4,#5,#6,#7,#8){\!\!\spicb{#2(26.25,15)(15,256,346)%
 #3(26.25,15)(15,-14,76) #4(30,30)(15,30) #1(18.75,15)(15,104,284)%
 #7(22.5,0)(15,30) #6(30,30)(26.25,15) #8(26.25,15)(22.5,0)%
 #5(25.25,15)(39.8,11.4)}\!\!}
\def\TopfVB(#1,#2,#3,#4,#5,#6,#7){\picb{#2(30,15)(15,-120,120)%
 #6(30,15)(15,120,180) #5(30,15)(15,180,240) #1(15,15)(15,60,300)%
 #4(15,15)(15,-60,0) #3(15,15)(15,0,60) #7(30,15)(15,15)}}
\def\STopfVB(#1,#2,#3,#4,#5,#6,#7){\spicb{#2(30,15)(15,-120,120)%
 #6(30,15)(15,120,180) #5(30,15)(15,180,240) #1(15,15)(15,60,300)%
 #4(15,15)(15,-60,0) #3(15,15)(15,0,60) #7(30,15)(15,15)}}
\def\TopfVN(#1,#2,#3,#4,#5,#6,#7){\picb{#1(15,15)(15,90,270)%
 #2(30,15)(15,-90,90) #4(30,30)(15,30) #3(15,0)(30,0)%
 #5(15,0)(15,30) #6(30,30)(30,0) #7(15,30)(30,0)}} 
\def\STopfVN(#1,#2,#3,#4,#5,#6,#7){\spicb{#1(15,15)(15,90,270)%
 #2(30,15)(15,-90,90) #4(30,30)(15,30) #3(15,0)(30,0)%
 #5(15,0)(15,30) #6(30,30)(30,0) #7(15,30)(30,0)}} 
\def\TopfVU(#1,#2,#3,#4,#5,#6,#7){\pic{#3(15,15)(15,0,90)%
 #2(15,15)(15,90,180) #4(15,15)(15,180,270) #1(15,15)(15,270,360)%
 #6(0,15)(15,30) #7(15,0)(0,15) #5(30,15)(15,0)}}
\def\STopfVU(#1,#2,#3,#4,#5,#6,#7){\spic{#3(15,15)(15,0,90)%
 #2(15,15)(15,90,180) #4(15,15)(15,180,270) #1(15,15)(15,270,360)%
 #6(0,15)(15,30) #7(15,0)(0,15) #5(30,15)(15,0)}}
\def\TopfVT(#1,#2,#3,#4,#5,#6){\pic{#1(15,15)(15,90,210)%
 #2(15,15)(15,210,330) #3(15,15)(15,-30,90) #4(2,7.5)(15,30)%
 #6(28,7.5)(2,7.5) #5(15,30)(28,7.5)}}
\def\STopfVT(#1,#2,#3,#4,#5,#6){\spic{#1(15,15)(15,90,210)%
 #2(15,15)(15,210,330) #3(15,15)(15,-30,90) #4(2,7.5)(15,30)%
 #6(28,7.5)(2,7.5) #5(15,30)(28,7.5)}}
\def\STopEVa(#1,#2){\spicb{#1(15,15)(15,90,270)%
 #1(30,15)(15,-90,90) #2(30,30)(15,30) #2(15,0)(30,0) #2(15,0)(15,30)%
 #2(30,30)(30,0) #2(15,15)(30,15) #2(22.5,15)(22.5,0)}}
\def\STopEVb(#1,#2){\spicb{#1(15,15)(15,90,270)%
 #1(30,15)(15,-90,90) #2(30,30)(15,30) #2(15,0)(30,0) #2(15,0)(15,30)%
 #2(30,30)(30,0) #2(15,10)(30,10) #2(15,20)(30,20)}}
\def\STopEVc(#1,#2){\spicb{#1(15,15)(15,90,270)%
 #1(30,15)(15,-90,90) #2(30,30)(15,30) #2(15,0)(30,0) #2(15,0)(15,30)%
 #2(30,30)(30,0) #2(15,10)(30,20) #2(15,20)(19.5,17) #2(25.5,13)(30,10)}}
\def\STopEVd(#1,#2){\spicb{#1(15,15)(15,90,270)%
 #1(30,15)(15,-90,90) #2(30,30)(15,30) #2(15,0)(30,0) #2(15,0)(15,30)%
 #2(30,30)(30,0) #2(15,11.25)(30,11.25) #2(22.5,9)(22.5,0)%
 #1(15,11.25)(7.5,15,90)}}
\def\STopEVe(#1,#2){\spic{#1(15,15)(15,0,90)%
 #1(15,15)(15,90,180) #1(15,15)(15,180,270) #1(15,15)(15,270,360)%
 #2(0,15)(15,30) #2(10,15)(10,1) #2(30,15)(0,15) #2(20,15)(20,1)}}
\def\STopEVf(#1,#2){\spic{#1(15,15)(15,0,90)%
 #1(15,15)(15,90,180) #1(15,15)(15,180,270) #1(15,15)(15,270,360)%
 #2(0,15)(15,30) #2(15,6)(24,15) #2(30,15)(0,15) #2(15,15)(15,0)}}
\def\STopEVg(#1,#2){\spicb{#1(15,15)(15,90,270)%
 #1(30,15)(15,-90,90) #2(30,30)(15,30) #2(15,0)(30,0) #2(15,0)(15,30)%
 #2(30,30)(30,0) #2(15,15)(45,15)}}
\def\STopEVh(#1,#2){\!\!\spicb{#1(26.25,15)(15,256,346)%
 #1(26.25,15)(15,-14,76) #2(30,30)(15,30) #1(18.75,15)(15,104,284)%
 #2(22.5,0)(15,30) #2(30,30)(26.25,15) #2(26.25,15)(22.5,0)%
 #2(25.25,15)(39.8,11.4) #2(19.75,15)(5.2,11.4)}\!\!}
\def\STopEVi(#1,#2){\spic{#1(15,15)(15,0,90)%
 #1(15,15)(15,90,180) #1(15,15)(15,180,270) #1(15,15)(15,270,360)%
 #2(15,15)(15,30) #2(0,15)(30,15) #2(15,0)(10,15) #2(15,0)(20,15)}}
\def\STopEVj(#1,#2){\spicb{#1(15,15)(15,90,270)%
 #1(30,15)(15,-90,90) #2(30,30)(15,30) #2(15,0)(30,0) #2(15,0)(15,20)%
 #2(15,20)(22.5,30) #2(22.5,30)(30,20) #2(30,20)(30,0)%
 #2(15,10)(30,20) #2(15,20)(19.5,17) #2(25.5,13)(30,10)}}
\def\STopEVk(#1,#2){\spicb{#1(15,15)(15,90,270)%
 #1(30,15)(15,-90,90) #2(30,30)(15,30) #2(15,0)(30,0) #2(15,0)(15,30)%
 #2(30,0)(30,30) #2(15,0)(30,24) #2(20,30)(24.5,19.2) #2(26.5,14.4)(30,6)}}
\def\STopEVl(#1,#2){\!\!\spicb{#1(26.25,15)(15,256,346)%
 #1(26.25,15)(15,-14,76) #2(30,30)(15,30) #1(18.75,15)(15,104,284)%
 #2(22.5,0)(5.2,11.4) #2(30,30)(26.25,15) #2(26.25,15)(22.5,0)%
 #2(26.25,15)(39.8,11.4) #2(15,30)(5.2,11.4) }\!\!}
\def\STopEVm(#1,#2){\spicb{#1(30,15)(15,-120,120) #1(30,15)(15,120,240)%
 #1(15,15)(15,60,300) #1(15,15)(15,-60,60) #2(22.5,15)(30,15)%
 #2(22.5,15)(17,22.5) #2(22.5,15)(17,7.5) }}
\def\STopEVn(#1,#2){\spicb{#1(15,15)(15,90,270)%
 #1(30,15)(15,-90,90) #2(30,30)(15,30) #2(15,0)(30,0)%
 #2(15,0)(15,30) #2(30,30)(30,0) #2(15,30)(30,0) #2(30,15)(45,15)}}
\def\STopEVo(#1,#2){\spicb{#1(30,15)(15,-120,120) #1(30,15)(15,120,240)%
 #1(15,15)(15,60,300) #1(15,15)(15,-60,60)%
 #2(16,20.5)(29,20.5) #2(16,9.5)(29,9.5) }}
\def\STopEVp(#1,#2){\spic{#1(15,15)(15,0,90)%
 #1(15,15)(15,90,180) #1(15,15)(15,180,270) #1(15,15)(15,270,360)%
 #2(0,15)(15,30) #2(0,15)(30,15) #2(15,0)(10,15) #2(15,0)(20,15)}}
\def\STopEVq(#1,#2){\spic{#1(15,15)(15,0,90)%
 #1(15,15)(15,90,180) #1(15,15)(15,180,270) #1(15,15)(15,270,360)%
 #2(0,15)(15,30) #1(19,15)(4,0,360) #2(30,15)(23,15) #2(15,15)(0,15)%
 #2(15,15)(15,0)}}
\def\STopEVr(#1,#2){\spic{#1(15,15)(15,0,90)%
 #1(15,15)(15,90,180) #1(15,15)(15,180,270) #1(15,15)(15,270,360)%
 #2(0,15)(15,30) #1(26,15)(4,0,360) #2(22,15)(0,15)%
 #2(15,15)(15,0)}}
\def\STopEVs(#1,#2){\spic{#1(15,15)(15,0,90)%
 #1(15,15)(15,90,180) #1(15,15)(15,180,270) #1(15,15)(15,270,360)%
 #2(0.75,10.35)(29.25,10.35) #2(15,0)(15,10.35)%
 #2(0.75,10.35)(6.2,27.1) #2(29.25,10.35)(23.8,27.1)}}
\def\STopEVt(#1,#2){\spic{#1(15,15)(15,0,90)%
 #1(15,15)(15,90,180) #1(15,15)(15,180,270) #1(15,15)(15,270,360)%
 #2(0,15)(15,30) #1(15,4)(4,0,360) #2(30,15)(0,15) #2(15,15)(15,8)}}
\def\STopEVu(#1,#2){\spic{#1(15,15)(15,0,90)%
 #1(15,15)(15,90,180) #1(15,15)(15,180,270) #1(15,15)(15,270,360)%
 #2(0,15)(15,30) #2(15,7.5)(30,15) #2(30,15)(0,15) #2(15,15)(15,0)}}
\def\STopEVv(#1,#2){\spic{#1(15,15)(15,0,90)%
 #1(15,15)(15,90,180) #1(15,15)(15,180,270) #1(15,15)(15,270,360)%
 #2(0,15)(30,15) #2(0,15)(15,7.5) #2(15,0)(15,15) #2(15,0)(22.5,15)}}
\def\STopEVw(#1,#2){\spicb{#1(30,15)(15,-120,120) #1(30,15)(15,120,240)%
 #1(15,15)(15,60,300) #1(15,15)(15,-60,60) #2(0,15)(15,15) #2(30,15)(45,15)}}
\def\STopEVx(#1,#2){\spicb{#1(15,15)(15,90,270)%
 #1(30,15)(15,-90,90) #2(30,30)(15,30) #2(15,0)(30,0) #2(15,0)(15,15)%
 #2(15,0)(30,15) #2(30,0)(30,15) #2(15,15)(20.5,9.5) #2(24.5,5.5)(30,0)%
 #2(15,15)(22.5,22.5) #2(30,15)(22.5,22.5) #2(22.5,30)(22.5,22.5)}}
\def\STopEVy(#1,#2){\spicb{#1(15,15)(15,90,270)%
 #1(30,15)(15,-90,90) #2(30,30)(15,30) #2(15,0)(30,0) #2(15,0)(15,30)%
 #2(30,0)(30,30) #2(15,0)(30,24) #2(15,30)(22.375,18.2) #2(26.375,11.8)(30,6)}}
\def\STopEVz(#1,#2){\spic{#1(15,15)(15,0,90)%
 #1(15,15)(15,90,180) #1(15,15)(15,180,270) #1(15,15)(15,270,360)%
 #2(15,0)(0.75,10.35) #2(15,0)(29.25,10.35)%
 #2(0.75,10.35)(6.2,27.1) #2(29.25,10.35)(23.8,27.1)}}
\def\STopEVaa(#1,#2){\spic{#1(15,15)(15,0,90)%
 #1(15,15)(15,90,180) #1(15,15)(15,180,270) #1(15,15)(15,270,360)%
 #2(0,15)(15,30) #2(15,30)(30,15) #2(30,15)(0,15) #2(15,15)(15,0)}}
\def\STopEVab(#1,#2){\spic{#1(15,15)(15,0,90)%
 #1(15,15)(15,90,180) #1(15,15)(15,180,270) #1(15,15)(15,270,360)%
 #2(15,0)(0.75,10.35) #2(15,0)(23.8,27.1)%
 #2(0.75,10.35)(6.2,27.1) #2(29.25,10.35)(23.8,27.1)}}
\def\STopEVac(#1,#2){\spic{#1(15,15)(15,0,90)%
 #1(15,15)(15,90,180) #1(15,15)(15,180,270) #1(15,15)(15,270,360)%
 #2(0,15)(15,30) #2(15,0)(30,15) #2(30,15)(0,15) #2(15,15)(15,0)}}
\def\STopEVad(#1,#2){\!\!\spicb{#1(26.25,15)(15,256,346)%
 #1(26.25,15)(15,-14,76) #2(30,30)(15,30) #1(18.75,15)(15,104,284)%
 #2(22.5,0)(15,30) #2(30,30)(26.25,15) #2(26.25,15)(22.5,0)%
 #1(33,13.2)(7,0,360)}\!\!}
\def\STopEVae(#1,#2){\spic{#1(15,15)(15,90,180)%
 #1(15,15)(15,180,270) #1(15,15)(15,270,360) #1(15,15)(15,0,90)%
 #2(15,15)(15,30) #2(15,15)(15,0) #2(0,15)(15,15) #2(30,15)(15,15)%
 #2(15,0)(30,15)}}
\def\STopEVaf(#1,#2){\spicb{#1(15,15)(15,90,270)%
 #1(30,15)(15,-90,90) #2(30,30)(15,30) #2(15,0)(30,0)%
 #2(15,0)(15,30) #2(30,30)(30,0) #2(15,30)(22.5,0) #2(22.5,0)(30,30)}}
\def\STopEVag(#1,#2){\spicb{#1(15,15)(15,90,270)%
 #1(30,15)(15,-90,90) #2(30,30)(15,30) #2(15,0)(30,0) #2(15,0)(15,15)%
 #2(15,15)(22.5,30) #2(22.5,30)(30,15) #2(30,15)(30,0) #2(15,15)(30,15)%
 #2(15,0)(30,15)}}
\def\STopEVah(#1,#2){\spic{#1(15,15)(15,0,90)%
 #1(15,15)(15,90,180) #1(15,15)(15,180,270) #1(15,15)(15,270,360)%
 #2(0,15)(15,30) #2(15,0)(0,15) #2(30,15)(15,0) #2(30,15)(15,30)}}
\def\STopEVai(#1,#2){\spicb{#1(30,15)(15,-120,120) #1(30,15)(15,120,240)%
 #1(15,15)(15,60,300) #1(15,15)(15,-60,60) #1(22.5,15)(7.5,0,360) }}
\def\TopoSTxt(#1,#2){\picb{#1(0,15)(20,15) #1(25,15)(45,15)%
 \GCirc(22.5,15){3}{0} \Text(22.5,20)[b]{{$\scriptstyle #2$}}}}
\def\TopoSB(#1,#2,#3){\picb{#1(0,15)(7.5,15) #2(22.5,15)(15,0,180)%
 #3(22.5,15)(15,180,360) #1(37.5,15)(45,15)}}
\def\STopoSB(#1,#2,#3){\spicb{#1(0,15)(7.5,15) #2(22.5,15)(15,0,180)%
 #3(22.5,15)(15,180,360) #1(37.5,15)(45,15)}}
\def\TopoST(#1,#2){\picb{#1(0,0)(22.5,0) #1(22.5,0)(45,0)%
 #2(22.5,15)(15,-90,270)}} 
\def\STopoST(#1,#2){\spicb{#1(0,0)(22.5,0) #1(22.5,0)(45,0)%
 #2(22.5,15)(15,-90,270)}} 
\def\ToptSiTxt(#1,#2){\picb{#1(0,15)(20,15) #1(25,15)(45,15)%
 \GBoxc(22.5,15)(6,6){0} \Text(22.5,20)[b]{{$\scriptstyle #2$}}}}
\def\ToptSM(#1,#2,#3,#4,#5,#6){\picb{#1(0,15)(7.5,15) #1(37.5,15)(45,15)%
 #2(22.5,15)(15,0,90) #3(22.5,15)(15,90,180) #4(22.5,15)(15,180,270)%
 #5(22.5,15)(15,270,360) #6(22.5,30)(22.5,0)}}
\def\ToptSAl(#1,#2,#3,#4,#5){\picb{#1(0,15)(7.5,15) #1(37.5,15)(45,15)%
 #2(22.5,15)(15,0,90) #3(22.5,15)(15,90,180) #4(22.5,15)(15,180,360)%
 #5(7.5,30)(15,270,360)}}
\def\ToptSAr(#1,#2,#3,#4,#5){\picb{#1(0,15)(7.5,15) #1(37.5,15)(45,15)%
 #2(22.5,15)(15,0,90) #3(22.5,15)(15,90,180) #4(22.5,15)(15,180,360)%
 #5(37.5,30)(15,180,270)}}
\def\ToptSE(#1,#2,#3,#4,#5){\picb{#1(0,15)(7.5,15) #1(37.5,15)(45,15)%
 #3(15,15)(7.5,0,180) #4(15,15)(7.5,180,360)%
 #2(30,15)(7.5,0,180) #5(30,15)(7.5,180,360)}} 
\def\ToptSS(#1,#2,#3,#4){\picb{#1(0,15)(7.5,15) #1(37.5,15)(45,15)%
 #4(7.5,15)(37.5,15) #2(22.5,15)(15,0,180) #3(22.5,15)(15,180,360)}}
\def\SToptSM(#1,#2,#3,#4,#5,#6){\spicb{#1(0,15)(7.5,15) #1(37.5,15)(45,15)%
 #2(22.5,15)(15,0,90) #3(22.5,15)(15,90,180) #4(22.5,15)(15,180,270)%
 #5(22.5,15)(15,270,360) #6(22.5,30)(22.5,0)}}
\def\SToptSAl(#1,#2,#3,#4,#5){\spicb{#1(0,15)(7.5,15) #1(37.5,15)(45,15)%
 #2(22.5,15)(15,0,90) #3(22.5,15)(15,90,180) #4(22.5,15)(15,180,360)%
 #5(7.5,30)(15,270,360)}}
\def\SToptSAr(#1,#2,#3,#4,#5){\spicb{#1(0,15)(7.5,15) #1(37.5,15)(45,15)%
 #2(22.5,15)(15,0,90) #3(22.5,15)(15,90,180) #4(22.5,15)(15,180,360)%
 #5(37.5,30)(15,180,270)}}
\def\SToptSE(#1,#2,#3,#4,#5){\spicb{#1(0,15)(7.5,15) #1(37.5,15)(45,15)%
 #3(15,15)(7.5,0,180) #4(15,15)(7.5,180,360)%
 #2(30,15)(7.5,0,180) #5(30,15)(7.5,180,360)}} 
\def\SToptSS(#1,#2,#3,#4){\spicb{#1(0,15)(7.5,15) #1(37.5,15)(45,15)%
 #4(7.5,15)(37.5,15) #2(22.5,15)(15,0,180) #3(22.5,15)(15,180,360)}}
\def\ToptSrTxt(#1,#2){\picb{#1(0,15)(20,15) #1(25,15)(45,15)%
 \GBoxc(22.5,15)(6,6){1} \Text(22.5,20)[b]{{$\scriptstyle #2$}}}}
\def\ToptSBB(#1,#2,#3){\picb{#1(0,15)(7.5,15)  #1(37.5,15)(45,15)%
 #2(22.5,15)(15,0,90) #2(22.5,15)(15,90,180) #3(22.5,15)(15,180,360)%
 \GCirc(22.5,30){5}{0.75} \Text(22.5,30)[c]{$\scriptstyle 1$}}}
\def\SToptSBB(#1,#2,#3){\spicb{#1(0,15)(7.5,15)  #1(37.5,15)(45,15)%
 #2(22.5,15)(15,0,90) #2(22.5,15)(15,90,180) #3(22.5,15)(15,180,360)%
 \GCirc(22.5,30){5}{0.75} \Text(16,21)[c]{$\scriptscriptstyle 1$}}}
\def\ToptSBBTxt(#1,#2,#3,#4){\picb{#1(0,15)(7.5,15)  #1(37.5,15)(45,15)%
 #2(22.5,15)(15,0,90) #2(22.5,15)(15,90,180) #3(22.5,15)(15,180,360)%
 \GCirc(22.5,30){3}{0} \Text(22.5,25)[t]{{$\scriptstyle #4$}}}}
\def\ToptSTB(#1,#2){\picb{#1(0,0)(22.5,0) #1(22.5,0)(45,0)%
 #2(22.5,15)(15,-90,90) #2(22.5,15)(15,90,270) \GCirc(22.5,30){5}{0.75}%
 \Text(22.5,30)[c]{$\scriptstyle 1$}}} 
\def\SToptSTB(#1,#2){\spicb{#1(0,0)(22.5,0) #1(22.5,0)(45,0)%
 #2(22.5,15)(15,-90,90) #2(22.5,15)(15,90,270) \GCirc(22.5,30){5}{0.75}%
 \Text(16,21)[c]{$\scriptscriptstyle 1$}}} 
\def\ToptSTBTxt(#1,#2,#3){\picb{#1(0,0)(22.5,0) #1(22.5,0)(45,0)%
 #2(22.5,15)(15,-90,90) #2(22.5,15)(15,90,270) \GCirc(22.5,30){3}{0}%
 \Text(22.5,25)[t]{{$\scriptstyle #3$}}}} 
\def\TopoT(#1,#2,#3){\!\pic{#1(15,30)(15,15) #2(2,7.5)(15,15)%
 #3(28,7.5)(15,15) \GCirc(15,15){3}{0}}\!} 
\def\TopoTTxt(#1,#2,#3,#4){\!\pic{#1(15,30)(15,15) #2(2,7.5)(15,15)%
 #3(28,7.5)(15,15) \GCirc(15,15){3}{0}%
 \Text(19,17)[bl]{{$\scriptstyle #4$}}}\!}
\def\TopoTS(#1,#2,#3,#4,#5,#6){\!\pic{#1(15,30)(15,23) #2(2,7.5)(8,11)%
 #3(28,7.5)(22,11) #6(15,15)(8,210,330) #5(15,15)(8,90,210)%
 #4(15,15)(8,-30,90)}\!}
\def\TopoTAo(#1,#2,#3,#4,#5){\!\pic{#1(15,30)(15,23) #2(2,7.5)(15,7)%
 #3(28,7.5)(15,7) #4(15,15)(8,-90,90) #5(15,15)(8,90,270)}\!}
\def\TopoTAr(#1,#2,#3,#4,#5){\!\pic{#1(15,30)(8,19) #2(2,7.5)(8,19)%
 #3(28,7.5)(22,11) #4(15,15)(8,-45,135) #5(15,15)(8,135,315)}\!}
\def\TopoTAl(#1,#2,#3,#4,#5){\!\pic{#1(15,30)(23,19) #2(2,7.5)(8,11)%
 #3(28,7.5)(23,19) #4(15,15)(8,45,225) #5(15,15)(8,-135,45)}\!}

\def\ToptrFpt(#1,#2,#3,#4){\picb{#1(0,0)(15,15) #2(15,15)(30,0)%
 #3(15,15)(30,30) #4(0,30)(15,15)}}
\def\ToptrFpta(#1,#2,#3,#4,#5){\picb{#1(0,0)(15,0) #2(15,0)(30,0)%
 #3(15,30)(30,30) #4(0,30)(15,30) #5(15,30)(15,0)}}
\def\ToptrFptb(#1,#2,#3,#4,#5){\picb{#1(0,0)(7,15) #2(23,15)(30,0)%
 #3(23,15)(30,30) #4(0,30)(7,15) #5(7,15)(23,15)}}
\def\ToptrFptc(#1,#2,#3,#4,#5){\picb{#1(0,0)(15,0) #2(15,30)(30,0)%
 #3(15,0)(20,10) #3(25,20)(30,30) #4(0,30)(15,30) #5(15,30)(15,0)}}
\def\ToptSfull(#1,#2){\picb{#1(0,15)(20,15) #1(25,15)(45,15)%
 \GCirc(22.5,15){3}{0.5} \Text(22.5,20)[b]{{$\scriptstyle #2$}}}}
\def\TopoSBfull(#1,#2,#3,#4){\picb{#1(0,15)(7.5,15) #2(22.5,15)(15,0,180)%
 #3(22.5,15)(15,180,360) #1(37.5,15)(45,15) \GCirc(37.5,15){3}{0.5}%
 \Text(32.5,15)[r]{{$\scriptstyle #4$}}}}
\def\ToptSSamp(#1,#2,#3,#4,#5){\picb{#1(0,15)(7.5,15) #1(37.5,15)(45,15)%
 #4(7.5,15)(37.5,15) #2(22.5,15)(15,0,180) #3(22.5,15)(15,180,360)%
 \GBoxc(37.5,15)(6,6){0.5} \Text(33.5,13)[tr]{{$\scriptstyle #5$}}}}
\def\ToptSSfull(#1,#2,#3,#4,#5){\picb{#1(0,15)(7.5,15) #1(37.5,15)(45,15)%
 #4(7.5,15)(37.5,15) #2(22.5,15)(15,0,180) #3(22.5,15)(15,180,360)%
 \GCirc(37.5,15){3}{0.5} \Text(33.5,13)[tr]{{$\scriptstyle #5$}}}}
\def\ToptSAlfull(#1,#2,#3,#4,#5,#6,#7){\picb{#1(0,15)(7.5,15)%
 #1(37.5,15)(45,15) #2(22.5,15)(15,0,90) #3(22.5,15)(15,90,180)%
 #4(22.5,15)(15,180,360) #5(7.5,30)(15,270,360)%
 \GCirc(37.5,15){3}{0.5} \GCirc(22.5,30){3}{0.5}%
 \Text(33.5,13)[tr]{{$\scriptstyle #6$}}%
 \Text(24.5,26)[tl]{{$\scriptstyle #7$}}}}
\def\ToptrFptamp(#1,#2,#3,#4){\picb{#1(4,4)(15,15) #2(15,15)(26,4)%
 #3(15,15)(26,26) #4(4,26)(15,15) \GBoxc(15,15)(6,6){0.5}}}
\def\ToptrFptfull(#1,#2,#3,#4){\picb{#1(4,4)(15,15) #2(15,15)(26,4)%
 #3(15,15)(26,26) #4(4,26)(15,15) \GCirc(15,15){3}{0.5}}}
\def\ToptrFptafull(#1,#2,#3,#4,#5){\picb{#1(4,4)(15,8) #2(15,8)(26,4)%
 #3(15,22)(26,26) #4(4,26)(15,22) #5(15,22)(15,8)%
 \GCirc(15,8){3}{0.5} \GCirc(15,22){3}{0.5}}}
\def\ToptrFptbfull(#1,#2,#3,#4,#5){\picb{#1(4,4)(8,15) #2(22,15)(26,4)%
 #3(22,15)(26,26) #4(4,26)(8,15) #5(8,15)(22,15)%
 \GCirc(8,15){3}{0.5} \GCirc(22,15){3}{0.5}}}
\def\ToptrFptcfull(#1,#2,#3,#4,#5){\picb{#1(4,4)(15,8) #2(15,22)(26,4)%
 #3(15,8)(18,12.9) #3(21,17.8)(26,26) #4(4,26)(15,22) #5(15,22)(15,8)%
 \GCirc(15,8){3}{0.5} \GCirc(15,22){3}{0.5}}}

\def\ToptVSblob(#1,#2,#3){\pic{#1(15,15)(15,0,180) #2(15,15)(15,180,360)%
 #3(30,15)(0,15) \GCirc(30,15){3}{0}}\;}
\def\ToptVSblobTxt(#1,#2,#3,#4){\pic{#1(15,15)(15,0,180) #2(15,15)(15,180,360)%
 #3(30,15)(0,15) \GCirc(30,15){3}{0} \Text(26,13)[tr]{{$\scriptstyle #4$}}}\;}
\def\ToprVBamp(#1,#2,#3,#4){\picb{#1(30,15)(15,-120,120) #2(30,15)(15,120,240)%
 #3(15,15)(15,60,300) #4(15,15)(15,-60,60) \GBoxc(22.5,28)(6,6){0}}}

\def\ToptVSblobsh(#1,#2,#3){\pic{#1(15,15)(15,0,180) #2(15,15)(15,180,360)%
 #3(30,15)(0,15) \GCirc(30,15){3}{0.5}}\;}
\def\SToptVSblobsh(#1,#2,#3){\spic{#1(15,15)(15,0,180) #2(15,15)(15,180,360)%
 #3(30,15)(0,15) \GCirc(30,15){3}{0.5}}\;}
\def\ToprVBampsh(#1,#2,#3,#4){\picb{#1(30,15)(15,-120,120)%
 #2(30,15)(15,120,240)%
 #3(15,15)(15,60,300) #4(15,15)(15,-60,60) \GBoxc(22.5,28)(6,6){0.5}}}
\def\ToprVVblob(#1,#2,#3,#4,#5){\!\!\picb{#2(26.25,15)(15,256,76)%
 #3(30,30)(15,30) #1(18.75,15)(15,104,284) #4(15,30)(22.5,0)%
 #5(30,30)(22.5,0) \GCirc(29,28){3}{0}}\!\!}
\def\ToprVVblobbsh(#1,#2,#3,#4,#5){\!\!\picb{#2(26.25,15)(15,256,76)%
 #3(30,30)(15,30) #1(18.75,15)(15,104,284) #4(15,30)(22.5,0)%
 #5(30,30)(22.5,0) \GCirc(29,28){3}{0.5} \GCirc(16,28){3}{0.5}}\!\!}
\def\SToprVVblobbsh(#1,#2,#3,#4,#5){\!\!\spicb{#2(26.25,15)(15,256,76)%
 #3(30,30)(15,30) #1(18.75,15)(15,104,284) #4(15,30)(22.5,0)%
 #5(30,30)(22.5,0) \GCirc(29,28){3}{0.5} \GCirc(16,28){3}{0.5}}\!\!}
\def\ToprVBblob(#1,#2,#3,#4){\picb{#1(30,15)(15,-120,120)%
 #2(30,15)(15,120,240)%
 #3(15,15)(15,60,300) #4(15,15)(15,-60,60) \GCirc(22.5,28){3}{0}}}
\def\ToprVBblobsh(#1,#2,#3,#4){\picb{#1(30,15)(15,-120,120)%
 #2(30,15)(15,120,240)%
 #3(15,15)(15,60,300) #4(15,15)(15,-60,60) \GCirc(22.5,28){3}{0.5}}}
\def\SToprVBblobsh(#1,#2,#3,#4){\spicb{#1(30,15)(15,-120,120)%
 #2(30,15)(15,120,240)%
 #3(15,15)(15,60,300) #4(15,15)(15,-60,60) \GCirc(22.5,28){3}{0.5}}}
\def\ToptrFptblob(#1,#2,#3,#4){\picb{#1(4,4)(15,15) #2(15,15)(26,4)%
 #3(15,15)(26,26) #4(4,26)(15,15) \GCirc(15,15){3}{0} }}

\def\SDgeneric(#1,#2,#3){\!\pic{#1(0,15)(15,15) #2(2,7.5)(15,15)%
 #3(2,22.5)(15,15) \GCirc(15,15){5}{0.75}%
 \Text(12,8)[c]{$\scriptstyle \ddots$}%
 \Text(8,2)[t]{{$\scriptstyle n$}}}\!} 
\def\SDoneLHS(#1){\pic{#1(0,15)(10,15) \GCirc(15,15){5}{0.5}}}
\def\SDoneA(#1,#2,#3){\picb{#1(0,15)(7.5,15) #2(22.5,15)(15,0,180)%
 #3(22.5,15)(15,180,360)}}
\def\SDoneB(#1,#2,#3,#4){\picb{#1(0,15)(7.5,15) #4(7.5,15)(37.5,15)%
 #2(22.5,15)(15,0,180) #3(22.5,15)(15,180,360) \GCirc(37.5,15){3}{0.5}}}
\def\SSDoneA(#1,#2,#3){\spicb{#1(0,15)(7.5,15) #2(22.5,15)(15,0,180)%
 #3(22.5,15)(15,180,360)}}
\def\SSDoneB(#1,#2,#3,#4){\spicb{#1(0,15)(7.5,15) #4(7.5,15)(37.5,15)%
 #2(22.5,15)(15,0,180) #3(22.5,15)(15,180,360) \GCirc(37.5,15){3}{0.5}}}
\def\SDtwoLHS(#1){\pic{#1(0,15)(10,15) #1(20,15)(30,15)%
 \GCirc(15,15){5}{0.5}}}
\def\SDtwoA(#1,#2,#3){\picb{#1(0,15)(7.5,15) #2(22.5,15)(15,0,180)%
 #3(22.5,15)(15,180,360) #1(37.5,15)(45,15) \GCirc(37.5,15){3}{0.5}}}
\def\SDtwoB(#1,#2,#3,#4){\!\picb{#1(0,0.5)(22.5,0)%
 #2(45,0.5)(22.5,0) #3(22.5,15)(15,-90,90) #4(22.5,15)(15,90,270)}}
\def\SDtwoC(#1,#2,#3,#4,#5){\picb{#1(0,15)(7.5,15) #1(37.5,15)(45,15)%
 #2(22.5,15)(15,0,90) #3(22.5,15)(15,90,180) #4(22.5,15)(15,180,360)%
 #5(7.5,30)(15,270,360) \GCirc(22.5,30){3}{0.5} \GCirc(37.5,15){3}{0.5}}}
\def\SDtwoD(#1,#2,#3,#4){\picb{#1(0,15)(7.5,15) #1(37.5,15)(45,15)%
 #4(7.5,15)(37.5,15) #2(22.5,15)(15,0,180) #3(22.5,15)(15,180,360)%
 \GCirc(37.5,15){3}{0.5}}}
\def\SSDtwoA(#1,#2,#3){\spicb{#1(0,15)(7.5,15) #2(22.5,15)(15,0,180)%
 #3(22.5,15)(15,180,360) #1(37.5,15)(45,15) \GCirc(37.5,15){3}{0.5}}}
\def\SSDtwoB(#1,#2,#3,#4){\!\spicb{#1(0,0.5)(22.5,0)%
 #2(45,0.5)(22.5,0) #3(22.5,15)(15,-90,90) #4(22.5,15)(15,90,270)}}
\def\SSDtwoC(#1,#2,#3,#4,#5){\spicb{#1(0,15)(7.5,15) #1(37.5,15)(45,15)%
 #2(22.5,15)(15,0,90) #3(22.5,15)(15,90,180) #4(22.5,15)(15,180,360)%
 #5(7.5,30)(15,270,360) \GCirc(22.5,30){3}{0.5} \GCirc(37.5,15){3}{0.5}}}
\def\SSDtwoD(#1,#2,#3,#4){\spicb{#1(0,15)(7.5,15) #1(37.5,15)(45,15)%
 #4(7.5,15)(37.5,15) #2(22.5,15)(15,0,180) #3(22.5,15)(15,180,360)%
 \GCirc(37.5,15){3}{0.5}}}
\def\SDthreeLHS(#1,#2,#3){\!\pic{#1(15,30)(15,15) #2(2,7.5)(15,15)%
 #3(28,7.5)(15,15) \GCirc(15,15){5}{0.5}%
 \Text(0,10)[b]{{$\scriptstyle 1$}} \Text(30,10)[b]{{$\scriptstyle 2$}}%
\Text(19,34)[t]{{$\scriptstyle 3$}}}\!} 
\def\SDthree(#1,#2,#3){\!\pic{#1(15,30)(15,15) #2(2,7.5)(15,15)%
 #3(28,7.5)(15,15)}\!}
\def\SSDthree(#1,#2,#3){\!\spic{#1(15,30)(15,15) #2(2,7.5)(15,15)%
 #3(28,7.5)(15,15)}\!}
\def\SSDthreeA(#1,#2,#3,#4,#5,#6){\!\spicb{#1(22.5,30)(22.5,40)%
 #2(-4.5,0)(9.5,7.5)%
 #3(48.5,0)(35.5,7.5) #6(22.5,15)(15,210,330) #5(22.5,15)(15,90,210)%
 #4(22.5,15)(15,-30,90) \GCirc(22.5,30){3}{0.5} \GCirc(35.5,7.5){3}{0.5}}\!}
\def\SSDthreeB(#1,#2,#3,#4,#5){\!\spicb{#1(22.5,40)(35.5,22.5)%
 #2(-4.5,0)(9.5,7.5)%
 #3(48.5,0)(35.5,22.5) #4(22.5,15)(15,45,225) #5(22.5,15)(15,-135,45)%
 \GCirc(35.5,22.5){3}{0.5}}\!}
\def\SSDthreeC(#1,#2,#3,#4,#5){\!\spicb{#1(22.5,30)(22.5,40)%
 #2(-4.5,0.5)(22.5,0)%
 #3(48.5,0.5)(22.5,0) #4(22.5,15)(15,-90,90) #5(22.5,15)(15,90,270)%
 \GCirc(22.5,30){3}{0.5}}\!}
\def\SSDthreeD(#1,#2,#3,#4,#5,#6,#7){\!\spicb{#1(22.5,30)(22.5,40)%
 #2(-4.5,0)(9.5,7.5)%
 #3(48.5,0)(35.5,7.5) #6(22.5,15)(15,210,330) #5(22.5,15)(15,90,210)%
 #4(22.5,15)(15,-30,90) #7(22.5,-5)(18,52,131) \GCirc(22.5,30){3}{0.5}%
 \GCirc(35.5,7.5){3}{0.5}}\!}
\def\SSDthreeE(#1,#2,#3,#4,#5,#6,#7,#8){\!\spicb{#1(22.5,30)(22.5,40)%
 #2(-4.5,0)(9.5,7.5)%
 #3(48.5,0)(35.5,7.5) #7(22.5,15)(15,210,330) #5(22.5,15)(15,90,150)%
 #4(22.5,15)(15,-30,90)  #6(22.5,15)(15,150,210) #8(7,15)(8,-80,80)%
 \GCirc(22.5,30){3}{0.5} \GCirc(35.5,7.5){3}{0.5} \GCirc(9.5,22.5){3}{0.5}}\!}
\def\SSDthreeF(#1,#2,#3,#4,#5,#6,#7,#8){\!\spicb{#1(22.5,30)(22.5,40)%
 #2(-4.5,0)(9.5,7.5) #3(48.5,0)(35.5,7.5)%
 #7(22.5,15)(15,210,330) #6(22.5,15)(15,90,210)%
 #4(22.5,15)(15,-30,30) #5(22.5,15)(15,30,90) #8(9.5,7.5)(35.5,22.5)%
 \GCirc(22.5,30){3}{0.5} \GCirc(35.5,22.5){3}{0.5} \GCirc(35.5,7.5){3}{0.5}}\!}
\def\SSDthreeG(#1,#2,#3,#4,#5,#6,#7){\!\spicb{#1(22.5,40)(35.5,22.5)%
 #2(-4.5,0)(9.5,7.5)%
 #3(48.5,0)(35.5,22.5) #4(22.5,15)(15,45,225) #5(22.5,15)(15,-135,-90)%
 #6(22.5,15)(15,-90,45)%
 #7(7,15)(8,-80,80) \GCirc(35.5,22.5){3}{0.5} \GCirc(9.5,22.5){3}{0.5}}\!}
\def\SSDthreeH(#1,#2,#3,#4,#5,#6){\!\spicb{#1(22.5,40)(35.5,22.5)%
 #2(-4.5,0)(9.5,7.5)%
 #3(48.5,0)(35.5,22.5) #4(22.5,15)(15,45,225) #5(22.5,15)(15,-135,45)%
 #6(9.5,7.5)(35.5,22.5) \GCirc(35.5,22.5){3}{0.5}}\!}
\def\SDfourLHS(#1,#2,#3,#4){\!\pic{#1(4,4)(15,15) #2(15,15)(26,4)%
 #3(15,15)(26,26) #4(4,26)(15,15) \GCirc(15,15){5}{0.5}%
 \Text(0,2)[b]{{$\scriptstyle 1$}}\Text(30,2)[b]{{$\scriptstyle 2$}}%
 \Text(30,28)[t]{{$\scriptstyle 3$}}\Text(0,28)[t]{{$\scriptstyle 4$}}%
}\!}
\def\SDfour(#1,#2,#3,#4){\!\pic{#1(4,4)(15,15) #2(15,15)(26,4)%
 #3(15,15)(26,26) #4(4,26)(15,15)}\!}
\def\SSDfour(#1,#2,#3,#4){\!\spic{#1(4,4)(15,15) #2(15,15)(26,4)%
 #3(15,15)(26,26) #4(4,26)(15,15)}\!}
\def\SSDfourA(#1,#2,#3,#4,#5,#6,#7,#8){\!\spicb{#1(0,-5)(12,4.5)%
 #2(45,-5)(33,4.5) #3(45,35)(33,25.5) #4(0,35)(12,25.5)%
 #5(22.5,15)(15,-135,-45) #6(22.5,15)(15,-45,45) #7(22.5,15)(15,45,135)%
 #8(22.5,15)(15,135,-135) \GCirc(33,4.5){3}{0.5}%
 \GCirc(33,25.5){3}{0.5} \GCirc(12,25.5){3}{0.5}}\!}
\def\SSDfourB(#1,#2,#3,#4,#5,#6,#7){\!\spicb{#1(0,-5)(12,4.5)%
 #2(45,-5)(33,4.5) #3(45,35)(22.5,30) #4(0,35)(22.5,30)%
 #5(22.5,15)(15,-135,-45) #6(22.5,15)(15,-45,90) #7(22.5,15)(15,90,-135)%
 \GCirc(33,4.5){3}{0.5} \GCirc(22.5,30){3}{0.5}}\!}
\def\SSDfourC(#1,#2,#3,#4,#5,#6){\!\spicb{#1(0,-5)(12,4.5)%
 #2(45,15)(33,25.5) #3(45,35)(33,25.5) #4(22.5,35)(33,25.5)%
 #5(22.5,15)(15,-135,45) #6(22.5,15)(15,45,-135) \GCirc(33,25.5){3}{0.5}}\!}
\def\SSDfourD(#1,#2,#3,#4,#5,#6,#7){\!\spicb{#1(0,-5)(22.5,0)%
 #2(45,-5)(22.5,0) #3(45,35)(33,25.5) #4(0,35)(12,25.5)%
 #5(22.5,15)(15,-90,45) #6(22.5,15)(15,45,135) #7(22.5,15)(15,135,-90)%
 \GCirc(33,25.5){3}{0.5} \GCirc(12,25.5){3}{0.5}}\!}
\def\SSDfourE(#1,#2,#3,#4,#5,#6){\!\spicb{#1(0,-5)(22.5,0)%
 #2(45,-5)(22.5,0) #3(45,35)(22.5,30) #4(0,35)(22.5,30)%
 #5(22.5,15)(15,-90,90) #6(22.5,15)(15,90,-90) \GCirc(22.5,30){3}{0.5}}\!}
\def\SDfiveLHS(#1,#2,#3,#4,#5){\!\pic{#1(15,15)(15,29) #2(15,15)(2.5,21)%
 #3(15,15)(7.5,3) #4(15,15)(22.5,3)  #5(15,15)(27.5,21)%
 \GCirc(15,15){5}{0.5}}\!}

\def\FullProp(#1){\pic{#1(5,15)(25,15) \Text(15,18)[b]{{\tiny (full)}}}}
\def\Prop(#1){\pic{#1(5,15)(25,15)}}
\def\PropPiProp(#1){\picb{#1(0,15)(15,15) #1(25,15)(40,15)
 \GCirc(20,15){5}{0.75} \Text(20,15)[c]{$\textstyle \pi$}}}
\def\PropFreeProp(#1){\picb{#1(0,15)(17.5,15) #1(22.5,15)(40,15)
 \Line(17.5,10)(22.5,20) \GCirc(20,15){2}{0} }}
\def\SPropPiProp(#1){\!\pic{#1(5,15)(15,15) #1(25,15)(35,15)
 \GCirc(20,15){5}{0.75} \Text(20,15)[c]{$\textstyle \pi$}}}
\def\SPropFreeProp(#1){\!\pic{#1(5,15)(17.5,15) #1(22.5,15)(35,15)
 \Line(17.5,10)(22.5,20) \GCirc(20,15){2}{0} }}
\def\FProp(#1){\picb{#1(0,15)(17.5,15) #1(27.5,15)(45,15)
 \GCirc(22.5,15){5}{0.75} }}
\def\FFProp(#1){\picb{#1(0,15)(17.5,15)}}
\def\PropPiPropPiProp(#1){\picc{#1(0,15)(15,15) #1(25,15)(35,15)%
 #1(45,15)(60,15) \GCirc(20,15){5}{0.75} \Text(20,15)[c]{$\textstyle \pi$}%
 \GCirc(40,15){5}{0.75} \Text(40,15)[c]{$\textstyle \pi$}}}
\def\SDvacO(#1,#2){\pic{#1(15,15)(15,-90,270)%
 \Line(13,-5)(17,5) \GCirc(15,0){2}{0} \Text(15,26)[t]{{\tiny #2}}}}
\def\SSDvacO(#1,#2){\spic{#1(15,15)(15,-90,270)%
 \Line(13,-5)(17,5) \GCirc(15,0){2}{0} \Text(15,26)[t]{{\tiny #2}}}}
\def\SDvacOpi(#1){\pic{#1(15,15)(15,91,89)%
 \GCirc(15,30){5}{0.75} \Text(15,30)[c]{$\textstyle \pi$}}}
\def\SSDvacOpi(#1){\spic{#1(15,15)(15,91,89)%
 \GCirc(15,30){5}{0.75} \Text(10.5,21)[c]{$\scriptstyle \pi$}}}
\def\ToprVMa(#1,#2,#3,#4,#5,#6){\spic{#3(15,15)(15,-30,90)%
 #1(15,15)(15,90,210)%
 #2(15,15)(15,210,330) #5(2,7.5)(15,15) #6(15,15)(15,30) #4(28,7.5)(15,15)%
 \GCirc(15,30){3}{0.5} \GCirc(28,7.5){3}{0.5}}}
\def\SToprVMa(#1,#2,#3,#4,#5,#6){\spic{#3(15,15)(15,-30,90)%
 #1(15,15)(15,90,210)%
 #2(15,15)(15,210,330) #5(2,7.5)(15,15) #6(15,15)(15,30) #4(28,7.5)(15,15)%
 \GCirc(15,30){3}{0.5} \GCirc(28,7.5){3}{0.5}}}
\def\ToprVMb(#1,#2,#3,#4,#5,#6,#7){\pic{#3(15,15)(15,-30,90)%
 #1(15,15)(15,90,210) #2(15,15)(15,210,330) #5(2,7.5)(15,15)%
 #6(15,15)(15,30) #4(28,7.5)(15,15)%
 \GCirc(15,30){3}{0.5} \Text(18,29)[bl]{#7}}}
\def\SToprVMb(#1,#2,#3,#4,#5,#6,#7){\spic{#3(15,15)(15,-30,90)%
 #1(15,15)(15,90,210) #2(15,15)(15,210,330) #5(2,7.5)(15,15)%
 #6(15,15)(15,30) #4(28,7.5)(15,15)%
 \GCirc(15,30){3}{0.5} \Text(18,29)[bl]{#7}}}
\def\ToprVVb(#1,#2,#3,#4,#5){\!\!\picb{#2(26.25,15)(15,256,76)%
 #3(30,30)(15,30) #1(18.75,15)(15,104,284) #4(15,30)(22.5,0)%
 #5(30,30)(22.5,0) \GCirc(22.5,0){3}{0.5}}\!\!}
\def\SToprVVb(#1,#2,#3,#4,#5){\!\!\spicb{#2(26.25,15)(15,256,76)%
 #3(30,30)(15,30) #1(18.75,15)(15,104,284) #4(15,30)(22.5,0)%
 #5(30,30)(22.5,0) \GCirc(22.5,0){3}{0.5}}\!\!}
\def\ToprVVc(#1,#2,#3,#4,#5,#6){\!\!\picb{#2(26.25,15)(15,256,76)%
 #3(30,30)(15,30) #1(18.75,15)(15,104,284) #4(15,30)(22.5,0)%
 #5(30,30)(22.5,0) \GCirc(22.5,0){3}{0.5} \Text(26,-1)[tl]{#6}}\!\!}
\def\TopfVXa(#1,#2,#3,#4,#5,#6,#7,#8,#9){\picb{#1(15,15)(15,90,270)%
 #2(30,15)(15,-90,90) #4(30,30)(15,30) #3(15,0)(30,0) #6(15,0)(15,15)%
 #5(15,15)(30,30) #8(15,30)(20,25) #8(25,20)(30,15) #7(30,15)(30,0)%
 #9(15,15)(30,15) \GCirc(15,0){3}{0.5} \GCirc(30,0){3}{0.5}%
 \GCirc(30,30){3}{0.5}}}
\def\TopfVHa(#1,#2,#3,#4,#5,#6,#7,#8,#9){\picb{#1(15,15)(15,90,270)%
 #2(30,15)(15,-90,90) #4(30,30)(15,30) #3(15,0)(30,0) #6(15,0)(15,15)%
 #5(15,15)(15,30) #8(30,30)(30,15) #7(30,15)(30,0) #9(15,15)(30,15)%
 \GCirc(30,0){3}{0.5} \GCirc(30,30){3}{0.5}}}
\def\TopfVHb(#1,#2,#3,#4,#5,#6,#7,#8,#9){\picb{#1(15,15)(15,90,270)%
 #2(30,15)(15,-90,90) #4(30,30)(15,30) #3(15,0)(30,0) #6(15,0)(15,15)%
 #5(15,15)(15,30) #8(30,30)(30,15) #7(30,15)(30,0) #9(15,15)(30,15)%
 \GCirc(15,0){3}{0.5} \GCirc(30,0){3}{0.5} \GCirc(30,30){3}{0.5}}}
\def\TopfVWa(#1,#2,#3,#4,#5,#6,#7,#8){\pic{#1(15,15)(15,90,180)%
 #3(15,15)(15,180,270) #2(15,15)(15,270,360) #4(15,15)(15,0,90)%
 #5(15,15)(15,30) #7(15,15)(15,0) #6(0,15)(15,15) #8(30,15)(15,15)
 \GCirc(15,0){3}{0.5} \GCirc(30,15){3}{0.5} \GCirc(15,30){3}{0.5}}\,}
\def\TopfVWb(#1,#2,#3,#4,#5,#6,#7,#8){\pic{#1(15,15)(15,90,180)%
 #3(15,15)(15,180,270) #2(15,15)(15,270,360) #4(15,15)(15,0,90)%
 #5(15,15)(15,30) #7(15,15)(15,0) #6(0,15)(15,15) #8(30,15)(15,15)
 \GCirc(15,15){3}{0.5} \GCirc(30,15){3}{0.5}}\,}
\def\TopfVWc(#1,#2,#3,#4,#5,#6,#7,#8){\pic{#1(15,15)(15,90,180)%
 #3(15,15)(15,180,270) #2(15,15)(15,270,360) #4(15,15)(15,0,90)%
 #5(15,15)(15,30) #7(15,15)(15,0) #6(0,15)(15,15) #8(30,15)(15,15)
 \GCirc(30,15){3}{0.5} \GCirc(15,30){3}{0.5}}\,}
\def\TopfVVa(#1,#2,#3,#4,#5,#6,#7,#8){\!\!\picb{#2(26.25,15)(15,256,346)%
 #3(26.25,15)(15,-14,76) #4(30,30)(15,30) #1(18.75,15)(15,104,284)%
 #7(22.5,0)(15,30) #6(30,30)(26.25,15) #8(26.25,15)(22.5,0)%
 #5(26.25,15)(39.8,11.4) \GCirc(26.25,15){3}{0.5}%
 \GCirc(39.8,11.4){3}{0.5}}\!}
\def\TopfVVb(#1,#2,#3,#4,#5,#6,#7,#8){\!\!\picb{#2(26.25,15)(15,256,346)%
 #3(26.25,15)(15,-14,76) #4(30,30)(15,30) #1(18.75,15)(15,104,284)%
 #7(22.5,0)(15,30) #6(30,30)(26.25,15) #8(26.25,15)(22.5,0)%
 #5(26.25,15)(39.8,11.4) \GCirc(15,30){3}{0.5} \GCirc(30,30){3}{0.5}%
 \GCirc(39.8,11.4){3}{0.5}}\!}
\def\TopfVVc(#1,#2,#3,#4,#5,#6,#7,#8){\!\!\picb{#2(26.25,15)(15,256,346)%
 #3(26.25,15)(15,-14,76) #4(30,30)(15,30) #1(18.75,15)(15,104,284)%
 #7(22.5,0)(15,30) #6(30,30)(26.25,15) #8(26.25,15)(22.5,0)%
 #5(26.25,15)(39.8,11.4) \GCirc(22.5,0){3}{0.5}}\!}
\def\TopfVVd(#1,#2,#3,#4,#5,#6,#7,#8){\!\!\picb{#2(26.25,15)(15,256,346)%
 #3(26.25,15)(15,-14,76) #4(30,30)(15,30) #1(18.75,15)(15,104,284)%
 #7(22.5,0)(15,30) #6(30,30)(26.25,15) #8(26.25,15)(22.5,0)%
 #5(26.25,15)(39.8,11.4) \GCirc(15,30){3}{0.5}}\!}
\def\TopfVVe(#1,#2,#3,#4,#5,#6,#7,#8){\!\!\picb{#2(26.25,15)(15,256,346)%
 #3(26.25,15)(15,-14,76) #4(30,30)(15,30) #1(18.75,15)(15,104,284)%
 #7(22.5,0)(15,30) #6(30,30)(26.25,15) #8(26.25,15)(22.5,0)%
 #5(26.25,15)(39.8,11.4) \GCirc(22.5,0){3}{0.5} \GCirc(39.8,11.4){3}{0.5}}\!}
\def\TopfVBa(#1,#2,#3,#4,#5,#6,#7){\picb{#2(30,15)(15,-120,120)%
 #6(30,15)(15,120,180) #5(30,15)(15,180,240) #1(15,15)(15,60,300)%
 #4(15,15)(15,-60,0) #3(15,15)(15,0,60) #7(30,15)(15,15)%
 \GCirc(22.5,2.5){3}{0.5} \GCirc(30,15){3}{0.5}}}
\def\TopfVBb(#1,#2,#3,#4,#5,#6,#7){\picb{#2(30,15)(15,-120,120)%
 #6(30,15)(15,120,180) #5(30,15)(15,180,240) #1(15,15)(15,60,300)%
 #4(15,15)(15,-60,0) #3(15,15)(15,0,60) #7(30,15)(15,15)%
 \GCirc(15,15){3}{0.5} \GCirc(30,15){3}{0.5}}}
\def\TopfVBc(#1,#2,#3,#4,#5,#6,#7){\picb{#2(30,15)(15,-120,120)%
 #6(30,15)(15,120,180) #5(30,15)(15,180,240) #1(15,15)(15,60,300)%
 #4(15,15)(15,-60,0) #3(15,15)(15,0,60) #7(30,15)(15,15)%
 \GCirc(22.5,2.5){3}{0.5}}}
\def\TopfVNa(#1,#2,#3,#4,#5,#6,#7){\picb{#1(15,15)(15,90,270)%
 #2(30,15)(15,-90,90) #4(30,30)(15,30) #3(15,0)(30,0)%
 #5(15,0)(15,30) #6(30,30)(30,0) #7(15,30)(30,0) \GCirc(30,30){3}{0.5}}} 
\def\TopfVNb(#1,#2,#3,#4,#5,#6,#7){\picb{#1(15,15)(15,90,270)%
 #2(30,15)(15,-90,90) #4(30,30)(15,30) #3(15,0)(30,0)%
 #5(15,0)(15,30) #6(30,30)(30,0) #7(15,30)(30,0)%
 \GCirc(15,30){3}{0.5} \GCirc(30,30){3}{0.5}}} 
\def\TopfVUa(#1,#2,#3,#4,#5,#6,#7){\pic{#3(15,15)(15,0,90)%
 #2(15,15)(15,90,180) #4(15,15)(15,180,270) #1(15,15)(15,270,360)%
 #6(0,15)(15,30) #7(15,0)(0,15) #5(30,15)(15,0) \GCirc(15,0){3}{0.5}}}
\def\TopfVTa(#1,#2,#3,#4,#5,#6){\pic{#1(15,15)(15,90,210)%
 #2(15,15)(15,210,330) #3(15,15)(15,-30,90) #4(2,7.5)(15,30)%
 #6(28,7.5)(2,7.5) #5(15,30)(28,7.5) \GCirc(15,30){3}{0.5}}}
\def\SDlatA(#1,#2,#3,#4,#5,#6){\!\!\picb{#2(26.25,15)(15.5,256,76)%
 #4(30,30)(15,30) #1(18.75,15)(15.5,104,284) #5(15,30)(22.5,0)%
 #6(30,30)(22.5,0) #3(14.3,18)(20,-70,33) \GCirc(22.5,0){3}{0.5}}\!\!}
\def\SDlatB(#1,#2,#3,#4,#5,#6,#7){\!\!\picb{#2(26.25,15)(15.5,256,76)%
 #3(22.5,30)(15,30) #4(30,30)(22.5,30) #1(18.75,15)(15.5,104,284)%
 #5(15,30)(22.5,0) #6(30,30)(22.5,0) #7(22.5,0)(22.5,30)%
 \GCirc(22.5,0){3}{0.5}}\!\!}
\def\ToprVVblobshTxt(#1,#2,#3,#4,#5,#6){\!\!\picb{#2(26.25,15)(15,256,76)%
 #3(30,30)(15,30) #1(18.75,15)(15,104,284) #4(15,30)(22.5,0)%
 #5(30,30)(22.5,0) \GCirc(29,28){3}{0.5}%
 \Text(31,28)[bl]{#6}}\!\!}
\def\SToprVVblobshTxt(#1,#2,#3,#4,#5,#6){\!\!\spicb{#2(26.25,15)(15,256,76)%
 #3(30,30)(15,30) #1(18.75,15)(15,104,284) #4(15,30)(22.5,0)%
 #5(30,30)(22.5,0) \GCirc(29,28){3}{0.5}%
 \Text(31,28)[bl]{#6}}\!\!}

\def\STopLA(#1,#2){\spicc{#1(15,15)(15,0,360) #1(45,15)(15,0,360)%
 #2(30,15)(60,15)}}
\def\STopLB(#1,#2){\;\;\spic{#2(0,15)(30,15) #2(7.5,28)(22.5,2)%
 #2(22.5,28)(7.5,2)%
 #1(15,30)(7.5,330,210) #1(2,7.5)(7.5,90,330) #1(28,7.5)(7.5,210,90)}}
\def\STopLa(#1,#2){\spicb{#1(15,15)(15,90,270)%
 #1(30,15)(15,-90,90) #2(30,30)(15,30) #2(15,0)(30,0) #2(22.5,0)(22.5,30)%
 #2(22.5,0)(15,30) #2(22.5,0)(30,30)}}
\def\STopLb(#1,#2){\!\!\spicb{#1(26.25,15)(15.5,256,76)%
 #2(30,30)(15,30) #1(18.75,15)(15.5,104,284) #2(15,30)(22.5,0)%
 #2(30,30)(22.5,0) #1(15,17.8)(19.3,292.8,39.1)}\!\!}
\def\STopLc(#1,#2){\spicb{#1(30,15)(15,-120,120)%
 #1(30,15)(15,120,240) #1(15,15)(15,60,300) #1(15,15)(15,-60,60)%
 #2(22.5,3)(22.5,27)}}
\def\STopLd{\spicc{\CArc(15,15)(15,0,360) \CArc(45,15)(15,0,360)
\Line(15,0)(15,30) \Line(15,15)(30,15)}}
\def\STopLe{\spicb{\CArc(15,15)(15,90,270) \CArc(30,15)(15,270,90) 
\Line(15,30)(30,30) \Line(15,0)(30,0) 
\Line(15,30)(22.5,0) \Line(30,30)(22.5,0) \CArc(30,36)(6,0,360)}}
\def\STopLf{\spicb{\CArc(15,15)(15,90,270) \CArc(30,15)(15,270,90) 
\Line(15,30)(30,30) \Line(15,0)(30,0) 
\Line(15,0)(22.5,30) \Line(30,0)(22.5,30) \CArc(22.5,36)(6,0,360)}}
\def\STopLg{\spicc{\CArc(15,15)(15,0,360) \CArc(45,15)(15,0,360) 
\Oval(15,15)(5,15)(0)}}
\def\STopLh{\spicc{\CArc(15,15)(15,0,360) \CArc(45,15)(15,0,360)
\Line(0,15)(60,15)}}
\def\STopLi{\spiccc{\CArc(15,15)(15,0,360) \CArc(45,15)(15,0,360) 
\CArc(75,15)(15,0,360) \Line(30,15)(60,15)}}
\def\STopLj{\spicb{\CArc(15,15)(15,0,360) \Line(0,15)(30,15) 
\CArc(40,5)(7,225,45) \CArc(40,25)(7,315,135) 
\Line(30,15)(35,30) \Line(30,15)(45,20) 
\Line(30,15)(45,10) \Line(30,15)(35,0)}}
\def\STopLk{\spic{\CArc(5,5)(7,135,315) \CArc(5,25)(7,45,225) 
\CArc(25,5)(7,225,45) \CArc(25,25)(7,315,135) 
\Line(10,0)(20,30) \Line(0,10)(30,20) 
\Line(0,20)(30,10) \Line(10,30)(20,0)}}

\def\TopoSsh(#1){\pib{#1(0,15)(10,15) #1(12.5,15)(22.5,15)%
 \GCirc(11.25,15){2.0}{0}}}
\def\SPropCircProp(#1,#2){\!\pib{#1(5,15)(15,15) #1(25,15)(35,15)
 \GCirc(20,15){5}{0.75} \Text(20,15)[c]{$\scriptstyle #2$}}}
\def\SPropBoxProp(#1,#2){\!\pib{#1(5,15)(15,15) #1(25,15)(35,15)
 \GBoxc(20,15)(9,9){0.75} \Text(20,15)[c]{$\scriptstyle #2$}}}
\def\TopoS(#1){\SPropCircProp(#1,1)}
\def\ToptSi(#1){\SPropCircProp(#1,2)}
\def\ToptSr(#1){\SPropBoxProp(#1,2)}

\begin{document}

\begin{titlepage}
\begin{flushright}
CERN-TH/2001-248\\
hep-ph/0109100\\
\end{flushright}
\begin{centering}
\vfill
 
{\bf A SIMPLE WAY TO GENERATE HIGH ORDER VACUUM GRAPHS} 

\vspace{0.8cm}

K. Kajantie$^{\rm a,}$\footnote{keijo.kajantie@helsinki.fi}, 
M. Laine$^{\rm b,}$\footnote{mikko.laine@cern.ch},  
Y. Schr\"oder$^{\rm a,}$\footnote{york.schroder@helsinki.fi}

\vspace{0.8cm}

{\em $^{\rm a}$%
Department of Physics,
P.O.Box 9, FIN-00014 University of Helsinki, Finland\\}

\vspace{0.3cm}

{\em $^{\rm b}$%
Theory Division, CERN, CH-1211 Geneva 23,
Switzerland\\}

\vspace*{0.8cm}
 
\end{centering}
 
\noindent
We describe an efficient practical procedure for enumerating
and regrouping vacuum Feynman graphs of a given order in perturbation 
theory. The method is based on a combination of Schwinger-Dyson 
equations and the two-particle-irreducible (``skeleton'') expansion. 
The regrouping leads to skeletons containing only free propagators, 
together with ``ring diagrams'' containing all the self-energy 
insertions. As a consequence, relatively few diagrams need to be drawn 
and integrations carried out at any single stage of the computation 
and, in low dimensions, overlapping ultraviolet/infrared 
subdivergences can be cleanly isolated.
As an illustration we enumerate the graphs contributing 
to the 4-loop free energy in QCD, explicitly in a continuum and 
more compactly in a lattice regularization.   
\vfill
\noindent
 

\vspace*{1cm}
 
\noindent
CERN-TH/2001-248\\
November 2001 

\vfill

\end{titlepage}


\section{Introduction}

There are many physics contexts where multiloop 
Feynman diagram computations are carried out. In QED one 
goes up to 4-loop level (for reviews see, e.g.,~\cite{qed}), 
because experiments are so precise. In particle physics phenomenology,
particularly QCD, one goes up to 4-loop level 
(see, e.g.,~\cite{RML}), because the coupling constant 
is not small. In studying critical phenomena 
in the simplest O($N$) condensed matter systems, 
one goes up to 5-loop level~(see, e.g.,~\cite{GLTC}), 
because the effective expansion parameter is not small.

Studies of QCD at a finite temperature $T$ are faced 
with a similar challenge. Indeed, the coupling constant
expansion converges even worse than at zero temperature, 
requiring at least $T\gg 10^3 \Lambda_\rmi{QCD}$
to make any sense at all~\cite{bn,adjoint}. 
So far, though, only resummed 3-loop level has been 
reached for the simplest physical observable, 
the free energy~\cite{az}, 
because a broken Lorentz symmetry makes the analysis much more 
complicated than in the cases mentioned above. In fact, 
even in principle only one more order is (partly) computable, 
and then the expansion breaks down completely~\cite{IR}.
Multiloop computations are not useless, though: the infrared
problems can be isolated to a simple three-dimensional (3d)
effective field theory~\cite{dr} and studied non-perturbatively
there~\cite{a0cond}, but to convert the results to physical
units from lattice regularization still necessitates 
a number of fixed-order perturbative 
computations~\cite{framework,contlatt,moore_a}.

As the loop order increases, so does the computational 
effort. The sheer enumeration of various diagrams and their
symmetry factors becomes non-trivial. The group theoretic
and Lorentz structures of single graphs are involved. 
Finally, the scalar integrals remaining are hard
to evaluate analytically. It is therefore clear that 
ideally, one would like to automatise the whole procedure 
(for a review of the current status see, e.g.,~\cite{HS}).

In this paper we concentrate on the first step of any multiloop 
computation, the enumeration of various Feynman diagrams. This 
step should be the easiest to automatise, since all one needs 
is a straightforward evaluation of Wick contractions. 
Indeed, various packages, such 
as {\tt FeynArts}~\cite{FA} and {\tt QGRAF}~\cite{QG}, 
are available for determining 
$n$-point functions in a given particle physics model. 

For vacuum graphs in condensed matter systems 
a similar approach is possible. For the quartic O($N$)
scalar model the combinatorics is not yet too hard, 
but variants thereof already require some work. 
Consequently, graphical algorithms have been 
developed at 4-loop order and beyond for a number
of simple models~\cite{Klmany}.

In many cases, though, a straightforward generation of the 
full set of diagrams of a given loop order may not be the
ideal way to go. In realistic theories there are very many graphs, 
and all integrals would have to be evaluated on the same footing. 
This is almost impossible, particularly if many different masses appear.

Here we wish to present what would seem to us 
to be a maximally manageable setup. All vacuum graphs are
generated, but they are cleanly separated into two groups:
one, of two-particle-irreducible (2PI) ``skeletons'' with 
free propagators; the other, of ``ring diagrams'' 
with various self-energy insertions (see also~\cite{ma}).
The self-energies, in turn, are directly obtained from lower order skeletons.
We find that this setup economises the generation of the 
various graphs quite significantly. We also point out that
in low dimensions, relevant for statistical physics applications,  
the integrations remaining are qualitatively different in the two sets. 

As an illustration of the setup, we enumerate 
the diagrams contributing to the 4-loop free energy 
of finite temperature QCD 
(as well as QED and the symmetric phases of  the electroweak theory
and scalar electrodynamics). 
We hope, though, that the setup 
may be applicable to some other cases as well. That is why
we wish to separate it from the evaluation of the 
integrals arising in the finite $T$ context~\cite{inpre}, 
specific for that physical situation.

Our plan is the following. 
We summarise our basic notation in~\se\ref{se:notation},
reorganise the standard skeleton expansion in~\se\ref{se:skeleton}, 
review the Schwinger-Dyson equations for $n$-point and
vacuum graphs in~\se\ref{se:sd}, and combine them with
the modified skeleton expansion to obtain a generating formula
for skeleton diagrams in~\se\ref{se:combine}. 
The corresponding results are given for a 
lattice regularization of a generic model in~\se\ref{se:lattice}.
As an illustration, 
we show the loop expansion for the free energy of QCD 
and related models in~\se\ref{se:continuum}. 
We discuss some basic properties of our setup and conclude
in~\se\ref{se:integrals}.


\section{Notation}
\la{se:notation}

Let us start by introducing a concise notation. 
While the method is valid for any theory, we explicitly 
give all equations for a generic $\varphi^3+\varphi^4$ model. 
Later on we discuss more specific examples within this 
class, in particular the QCD, as well as some extensions of this class.
The generic class also includes the electroweak sector
of the Standard Model, both in its symmetric and its
spontaneously broken phase.  

The partition function is defined as
\be
Z[J] =  \int{\cal D}\varphi\, e^{S[\varphi]+J\varphi},
\la{eq:ZJ}
\ee
where $S[\varphi]$ is the action, 
\ba
S[\varphi] &=& -\fr12 \varphi_i \Delta^{-1}_{ij} \varphi_j
+ \fr1{3!} \gamma_{ijk} \varphi_i \varphi_j \varphi_k
+ \fr1{4!} \gamma_{ijkl} \varphi_i \varphi_j \varphi_k \varphi_l \;,
\la{eq:generic}
\ea
and summations over
various indices, numbering (real scalar) fields and their 
internal and spacetime structures, are implied.   
Two comments are in order. First, we will for the moment
not display fermions explicitly. As far as vacuum graphs are
concerned, they do not introduce any complications
apart from the usual overall minus sign for each closed loop, 
and can thus be introduced only at the end~\cite{Cvi}. Second, 
one should notice that the sign conventions in~\eqs\nr{eq:ZJ}, 
\nr{eq:generic} are such that in the case of
Euclidean actions, $\gamma_{ijkl}$ is typically negative. 

For a theory with a broken symmetry, the 
inverse free propagator $\Delta^{-1}$ and 
the couplings $\gamma_{ijk...}$ are functions of the order parameter, 
but otherwise there are no essential complications.
We return to this point in~\se\ref{se:npoint}.

The partition function $Z[J]$ in \eq\nr{eq:ZJ} 
is the generating functional for full Green's functions,
$\Gamma_n^{{\rm full}} = \ld \d_J^n\, Z[J] \right|_{J=0}$.
As usual we define 
\be
W[J] = \ln\,Z[J],
\la{eq:WJ}
\ee
the generating functional of connected Green's functions,
$\Gamma_n^{{\rm conn}} = \ld \d_J^n\, W[J] \right|_{J=0}$.
Finally, one can define the effective action via 
\be
S_{\rm eff}[\phi]=W[J]-\phi J, \quad
\phi = \delta_J W[J],
\la{eq:Seff}
\ee
which generates one-particle-irreducible (1PI) Green's
functions, $\Gamma_n^{{\rm 1PI}} =  \d_\phi^{\,n}\! 
\ld S_{\rm eff}[\phi] \right|_{\phi=0}$. Note, in particular, 
that $\delta_\phi S_\rmi{eff}[\phi] = - J$.
The vacuum, or free energy $F$ 
(made dimensionless by a division 
with the temperature~$T$), can be obtained from 
any of the generating functionals as
\be
F = -\ln Z[0] = -W[0] = -S_\rmi{eff}[0].
\la{eq:F}
\ee

{}From the basic relations $\phi = \delta_J W[J]$, 
$\delta_\phi S_\rmi{eff}[\phi] = - J$ it follows that
\be
\delta_J^2 W[J] \,
\delta_\phi^2 S_\rmi{eff}[\phi] = -1.
\la{eq:rel}
\ee
Defining, as usual, the ``proper'' self-energy by 
\be
\delta_\phi^2 S_\rmi{eff}[\phi] \equiv -\Delta^{-1} + \Pi, 
\ee
we see from \eq\nr{eq:rel} that
$\delta_J^2 W[J]$ is the full propagator: 
\be
\delta_J^2 W[J] \equiv D[\phi] = \frac{1}{\Delta^{-1} - \Pi} \equiv   
\Delta + \Delta \Pi \Delta + \Delta \Pi \Delta \Pi \Delta + ... \;.
\la{eq:fullprop}
\ee
We shall use here the following notation for free and full propagators, 
the proper self-energy, as well as general 1PI vertices:
\ba
 \Delta & = & \Prop(\Lsc)\quad 
              \hspace*{0.52cm} (\mbox{free propagator}), \la{freeprop} \\
      D & = & \Prop(\TLsc) = \Prop(\Lsc) +\PropPiProp(\Lsc) 
+\PropPiPropPiProp(\Lsc) +\dots (\mbox{full propagator}), 
\la{eq:fullproppic} \\
    \Pi & = & \PropPiProp(\DLsc) \quad 
    \mbox{(proper self-energy, with legs ``amputated'')}, \\
 \Delta^{-1} & = & \PropFreeProp(\DLsc)
 \quad \mbox{(inverse free propagator, with legs amputated)}, \\
 \delta_\phi^n S_\rmi{eff} & = & 
 \SDgeneric(\DLsc,\DLsc,\DLsc) \quad \hspace*{0.65cm} 
 (\mbox{general amputated 1PI vertex})\;.  \la{1pivertex}
\ea


\section{Skeleton expansion with free propagators}
\la{se:skeleton}

We next review the skeleton expansion 
for the free energy~$F$~\cite{LW,CJT}, and modify it such 
that full propagators can be replaced with free propagators~\cite{ma}.
By a skeleton we mean a 2PI vacuum diagram: one that remains 
connected even if any two lines are cut.
The skeleton expansion has been used as 
the starting point also in~\cite{ma}.

It can be shown~\cite{LW,CJT} 
that the loop expansion for~\eq\nr{eq:F} can be written as
\newcommand{\Tr}{{\rm Tr}}
\ba \label{Fskel}
F[D] = \sum_{\rm i} \, c_\rmi{i} \(\Tr\ln D_\rmi{i}^{-1} +
\Tr\,\Pi_\rmi{i}[D] D_\rmi{i}\) -\Phi[D] \;,
\ea
where ${\rm i} = \{\mbox{bosons}, \mbox{fermions}\}$, 
$c_\rmi{boson} = 1/2$, and $c_\rmi{fermion} = - 1$.  
Here $\Phi[D]$ collects all 2PI vacuum diagrams.
The full propagators~$D_\rmi{i}$ 
are related to their corresponding self-energies
by $D^{-1}=\Delta^{-1}-\Pi$ 
(cf.~\eq\nr{eq:fullprop}),
where $\Delta$ are the free propagators. 
Both $F$, $\Pi$ and $\Phi$ can be regarded as functionals of the full 
propagators. The partition function has an 
extremal property, such that the variation
of $F$ with respect to any of the full propagators 
vanishes~\cite{LW,CJT,RS}, 
giving a relation between skeletons and self-energies:
\ba \label{KapuRel}
\d_{D_\rmi{i}}\, \Phi[D] \;=\; c_\rmi{i} \Pi[D] \;.
\ea
Here, we have introduced the implicit notation that whenever
a term is multiplied by $c_\rmi{i}$, the $\Pi$'s and $D$'s 
following it are assumed to carry the same subscript. 
Pictorially, \eq\nr{KapuRel} corresponds to getting 
a self-energy by ``cutting a propagator'' 
in all possible ways in the set of vacuum skeletons.
Hence, knowing the skeletons alone provides full information.

In \eqs\nr{Fskel}, \nr{KapuRel}, it is the full propagators $D$
which appear in the skeleton graphs and self-energies. We would
instead like to obtain skeletons with free propagators.
As a first step in this direction, we expand $D$ in terms of the 
self-energy insertions $\Pi[D]$, $D=\Delta\sum_{n\geq0}(\Pi\Delta)^n$, to get 
\ba
F &=& \sum_\rmi{i} \! c_\rmi{i}\,\Tr\lk\ln\Delta^{-1}+\sum_{n\ge2}\(1\sy-1n\)
\(\Pi \Delta \)^n\rk - \Phi\Bigl[ \Delta\sum_{n\geq0}(\Pi\Delta)^n \Bigr] \;.
\la{eq:Fexp}
\ea 
We then have to evaluate $\Pi[D]$.

To go forward more explicitly,
we restrict ourselves to 5-loop level here.
Let the subscript $n$ denote the loop order, 
and write $\Pi=\sum_{n\ge 1}\Pi_n$. 
It turns out that we need at most $\Pi_3$. 
In a straightforward way, we obtain
\ba
\Pi_1 & = & \Pi_1[\Delta] \equiv \Pi_1^\rmi{irr}[\Delta], \\
\Pi_2 & = & \Pi_2^\rmi{irr}[\Delta] +  
 \Bigl( \Pi_1^\rmi{irr}[\Delta + \Delta \Pi_1^\rmi{irr} \Delta] \Bigr)_2  \nn
 & \equiv &  \Pi_2^\rmi{irr}[\Delta] + \Pi^{\rmi{red}(1)}_{2}[\Delta], \\
\Pi_3 & = & \Pi_3^\rmi{irr}[\Delta] +  
            \Bigl( \Pi_2^\rmi{irr} 
            [\Delta + \Delta \Pi_1^\rmi{irr} \Delta] \Bigr)_3  + 
            \Bigl( \Pi_1^\rmi{irr} 
            [\Delta + \Delta \Pi \Delta +
     \Delta \Pi_1^\rmi{irr} \Delta \Pi_1^\rmi{irr} \Delta ]\Bigr)_3  \nn
 & \equiv & \Pi_3^\rmi{irr}[\Delta] +  
       \Pi^{\rmi{red}(1)}_{3}[\Delta] + 
       \Pi^{\rmi{red}(2)}_{3}[\Delta] \;,
\ea
where $\Pi_n^\rmi{irr}$ are $n$-loop 
1PI graphs, while $\Pi_n^{\rmi{red}(m)}$
are obtained by cutting $m$ lines in
a lower order $\Pi_n^\rmi{irr}[\Delta]$,  
and dressing them appropriately:
\ba
  \Pi^{\rmi{red}(1)}_{2}[\Delta] & = & 
        (\Delta \Pi_1^\rmi{irr} \Delta)_\rmi{j} 
        \delta_{\Delta_\rmi{j}}\Pi_1^\rmi{irr}[\Delta], \la{red1} \\  
  \Pi^{\rmi{red}(1)}_{3}[\Delta] & = & 
  (\Delta \Pi_1^\rmi{irr} \Delta)_\rmi{j} 
        \delta_{\Delta_\rmi{j}}\Pi_2^\rmi{irr}[\Delta] +  
  (\Delta \Pi_2 \Delta + 
   \Delta \Pi_1^\rmi{irr} \Delta \Pi_1^\rmi{irr} \Delta)_\rmi{j} 
        \delta_{\Delta_\rmi{j}}\Pi_1^\rmi{irr}[\Delta], \\     
  \Pi^{\rmi{red}(2)}_{3}[\Delta] & = & \fr12
  (\Delta \Pi_1^\rmi{irr} \Delta)_\rmi{j} 
  (\Delta \Pi_1^\rmi{irr} \Delta)_\rmi{k} 
   \delta_{\Delta_\rmi{j}}
   \delta_{\Delta_\rmi{k}}
   \Pi_1^\rmi{irr}[\Delta] \;. \la{red3}
\ea
For the explicit diagrammatic characteristics of $\Pi^{\rmi{red}(1)}_{2}$, 
see \se\ref{se:genericSE}.

It is easy now to unfold the loop expansion also 
for $\Phi[D] =\sum_{n_\ge 2}\Phi_n$, the last 
term in \eq\nr{eq:Fexp}.
Up to 5-loop level, we can write
\ba
  \Bigl( \Phi_2[D] \Bigr)_{n\le 5} & = & 
    \Bigl(\Phi_2[\Delta + \Delta (\Pi_1+\Pi_2+\Pi_3)\Delta \nn
 & & + 
           \Delta (\Pi_1+\Pi_2)\Delta (\Pi_1+\Pi_2)\Delta + 
           \Delta \Pi_1 \Delta \Pi_1 \Delta \Pi_1 \Delta ]\Bigr)_{n\le 5}\;, \\
  \Bigl( \Phi_3[D] \Bigr)_{n\le 5} & = & 
    \Bigl(\Phi_3[\Delta + \Delta (\Pi_1+\Pi_2)\Delta + 
           \Delta \Pi_1 \Delta \Pi_1 \Delta ]\Bigr)_{n\le 5}\;, \\
  \Bigl( \Phi_4[D] \Bigr)_{n\le 5} & = & 
    \Bigl(\Phi_4[\Delta + \Delta \Pi_1 \Delta ]\Bigr)_{n\le 5}\;, \\
  \Bigl( \Phi_5[D] \Bigr)_{n\le 5} & = & 
    \Phi_5[\Delta ]\;,
\ea
where the arguments are to be Taylor expanded, with  
first derivatives obeying~(cf.~the dia\-grammatic 
identity \eq\nr{KapuRel}, evaluated with free propagators)
\be
\delta_{\Delta_\rmi{i}} \Phi_n[\Delta] = 
 c_\rmi{i} \Pi_{n-1}^\rmi{irr} [\Delta]\;, \la{eq:irrdef} 
\ee
and higher ones bringing back reducible self-energies, 
defined in \eqs\nr{red1}--\nr{red3}.

Inserting these expansions into \eq\nr{eq:Fexp} we finally get, 
up to 5-loop level, 
\ba
-F \!\!\! & = & \!\!\! 
-\sum_\rmi{i} \! c_\rmi{i}\,\Tr\ln\Delta^{-1}
+\Phi_2[\Delta] \nn 
 \!\!\! & + &  \!\!\! 
\Phi_3[\Delta] +\sum_\rmi{i} \! c_\rmi{i}\,\Tr \biggl[
\fr12\(\Delta\Pi_{1}\)^2\biggr] 
\nn
 \!\!\! & + & \!\!\! 
\Phi_4[\Delta] +\sum_\rmi{i} \! c_\rmi{i}\,\Tr \biggl[
\fr13\(\Delta\Pi_{1}\)^3 
+\Delta\Pi_{1}\Delta \Bigl( \Pi^\rmi{irr}_{2}
+\fr12 \Pi^{\rmi{red}(1)}_{2} \Bigr) \biggr] \nn
 \!\!\! & + &  \!\!\! 
\Phi_5[\Delta] +\sum_\rmi{i} \! c_\rmi{i}\,\Tr \biggl[
\fr14\( \Delta\Pi_{1} \)^4 
+ \( \Delta\Pi_{1} \)^2 \Delta \Bigl( \Pi_2^\rmi{irr} 
+ \fr12 \Pi_2^{\rmi{red}(1)} \Bigr)
 \nn
 \!\!\! & + &  \!\!\! 
\fr12 \Delta \Pi_2^\rmi{irr} \Delta \Bigl( \Pi_2^\rmi{irr} 
+ \Pi_2^{\rmi{red}(1)}\Bigr)
+\Delta \Pi_1 \Delta \Bigl( \Pi_3^\rmi{irr}
+\fr12 \Pi_3^{\rmi{red}(1)}
+\fr13 \Pi_3^{\rmi{red}(2)} \Bigr)
\biggr], 
\la{eq:skeleton}
\ea
or, written diagrammatically (and denoting by $F_0$ the 
non-interacting result), 
\ba
-F &=& -F_0 +\Phi_2[\Delta] 
\nonumber \\[0ex]&&{}
+\( \Phi_3[\Delta] +\displaystyle\sum_\rmi{i} \! c_\rmi{i}\,\(  
\sy{}12\STTopoVRoo(\Asc)\) \) 
\nonumber \\[0ex]&&{}
+\( \Phi_4[\Delta] +\displaystyle\sum_\rmi{i} \! c_\rmi{i}\(
\sy{}13\STTopoVRooo(\Asc)\sm{+}\STTopoVRoi(\Asc)\sy+12\STTopoVRor(\Asc)\) \)
\nonumber \\[0ex]&&{}
+\( \Phi_5[\Delta] +\displaystyle\sum_\rmi{i} \! c_\rmi{i}\(
\sy{}14\STTopoVRoooo(\Asc)\sm{+}\STTopoVRooi(\Asc)
\sy+12\STTopoVRoor(\Asc) \right.\right. 
\nonumber \\[0ex]&&{}
\left.\left.
\sy+12\STTopoVRiie(\Asc)\sy+12\STTopoVRire(\Asc)
\sm{+}\STTopoVRiif(\Asc)
\sy+12\STTopoVRirf(\Asc)\sy+13\STTopoVRirrf(\Asc)
\) \). \la{graphic}
\ea
Here a circle with $n$ inside denotes $\Pi^\rmi{irr}_n$, 
a square~$\Pi^{\rmi{red}(1)}_n$, 
and a double square~$\Pi^{\rmi{red}(2)}_n$.
We will term the skeletons with free propagators, $\Phi_n[\Delta]$,  
{\em irreducible}. Note that the numerical factors in front of 
various types of ring diagrams do not appear to trivially follow from 
any simple symmetry argument (particularly 
in the case of reducible self-energy
insertions), but are best worked out explicitly via
the Taylor expansions we have described. 

\eq\nr{graphic} is the starting point of our setup.
It expresses the free energy in an economic way 
in terms of the irreducible skeletons $\Phi_n[\Delta]$:
either as direct contributions, or as self-energy insertions, 
obtained from the same skeletons 
via \eqs\nr{eq:irrdef}, \nr{red1}--\nr{red3}. 
We note that at $n$-loop level, one needs 
$\Phi_n[\Delta]$ but only $\Pi_{n-2}[\Delta]$, 
obtained from $\Phi_{n-1}[\Delta]$.


\section{Schwinger-Dyson equations with full propagators}
\la{se:sd}

Next, we need to generate the skeletons $\Phi_n[\Delta]$, 
needed in \se\ref{se:skeleton}. To do that, 
we first review briefly the general
setup of Schwinger-Dyson (SD) equations, 
converted to our notation. The SD equations will then play
a central role in our main result, \eq\nr{SDvac}, 
which is an explicit formula allowing for a systematic
generation of all skeletons $\Phi_n[\Delta]$ --- 
in principle to any order.
In this section, we follow closely
the very enjoyable presentation by Cvitanovi\'c~\cite{Cvi}. 

\subsection{General $n$-point functions}
\la{se:npoint}

The basic SD equation for 
the generating functional $Z[J]$ of 
full Green's functions, 
derives from the trivial
fact that the integral of a total derivative vanishes:
\ba
0 &=& \int{\cal D}\varphi\, \d_\varphi\, e^{S[\varphi]+J\varphi} 
\;\;=\;\; \( S'[\d_J]+J \) Z[J] \;.
\ea
For the generating functional of the connected Green's
functions, \eq\nr{eq:WJ}, one gets
\ba 
0 &=& S'[W'[J]+\d_J] + J \;.
\ea
Finally, for the effective action, \eq\nr{eq:Seff}, 
we use from \se\ref{se:notation} that 
$W'[J] = \phi$,  
$\delta_J = (\delta \phi/\delta J)\delta_\phi = 
W''[J]\delta_\phi = D[\phi]\delta_\phi $, 
and $J = -S_\rmi{eff}'[\phi]$ to obtain  
\ba \label{basicSD}
S_{\rm eff}'[\phi] \;\;=\;\; S'[\phi+D[\phi]\d_\phi]
\;.
\ea
Putting $\phi\to 0$ on the right-hand-side, 
this gives the SD equation for the 1-point function, 
while taking derivatives with respect to $\phi$ on both sides
of \eq\nr{basicSD} and putting $\phi\to 0$ only afterwards, 
generates SD equations for higher-point Green's functions, 
\ba
\Gamma_n^{\rm 1PI} &=& \ld \d_\phi^{\,n-1}\, 
S'[\phi+D[\phi]\d_\phi] \right|_{\phi=0}.
\la{eq:1PI}
\ea 
Here $D[\phi]$ is in \eq\nr{eq:fullprop}, and we note that
\be
\d_\phi D[\phi] = D[\phi]\, 
\Bigl( \delta^3_\phi S_\rmi{eff}[\phi]\Bigr) \, D[\phi].
\la{eq:dD}
\ee

A note may be in order here, 
concerning theories with spontaneously 
broken symmetries. In that case, $\phi$ corresponds to the 
fluctuating field around some reference value $v$, 
typically $v \equiv \langle \varphi \rangle$. 
The quantity we should ultimately be 
computing is the free energy density as a function of $v$, 
i.e.\ the effective potential $V(v) = F/(\mbox{volume})$.
Then everything goes as before: we still put $\phi\to 0$
in the equations above after differentiation, 
while the condensate $v$ appears 
as a parameter in the free propagators as well as
in the cubic and quartic 
couplings in \eq\nr{eq:generic} (the term $J\varphi$ 
linear in $\varphi$ in \eq\nr{eq:ZJ}
need not be changed~\cite{effpot}). 
The graphs also remain the same: only 1PI graphs, 
generated by the loop expansion in~\eq\nr{graphic}, 
are to be included~\cite{effpot}. Tadpole type graphs
often associated with broken symmetries
would only be generated if we want to re-expand the 
value of $V(v)$ at the broken minimum in a strict loop expansion:
writing $V = \sum_{n\ge 0} V_n$, $v = \sum_{n\ge 0} v_n$, 
such that $V_0'(v_0)=0$, implies
\ba
\left. V(v) \right|_{V'(v) = 0}  & = &  
V_0(v_0) + V_1(v_0)+ 
\biggl[V_2 - \fr12 \frac{(V_1')^2}{V_0''} \biggr]_{v = v_0} \nn
&  + & 
\biggl[ 
V_3 - \frac{V_1'V_2'}{V_0''} + 
\fr12 \frac{(V_1')^2V_1''}{(V_0'')^2} - 
\fr16 \frac{(V_1')^3 V_0'''}{(V_0'')^3}
\biggr]_{v = v_0}
\nn
& + & 
\biggl[
V_4 -
\fr12 \frac{(V_2')^2 + 2 V_1' V_3'}{V_0''} + 
\fr12 \frac{2 V_1'V_2'V_1'' + (V_1')^2 V_2''}{(V_0'')^2} \nn
& & -
\fr16 \frac{3 (V_1')^2 (V_1'')^2+3(V_1')^2 V_2' V_0''' +
(V_1')^3 V_1'''}{(V_0'')^3} \nn 
& & +
\frac{1}{24}\frac{12 (V_1')^3 V_1'' V_0'''+(V_1')^4 V_0''''}{(V_0'')^4}-
\fr18 \frac{(V_1')^4 (V_0''')^2}{(V_0'')^5}
\biggr]_{v = v_0}
+ ... \;, \la{eq:tadpoles}
\ea 
where the latter terms inside the square brackets 
correspond to various tadpole graphs, with obvious
notation: $1/V_0''$ is the free propagator of the Higgs particle with
a vanishing momentum, $V_1'$ ($V_1''$) is a 1-loop diagram 
with one leg (two legs), $V_0'''$ is a three-vertex, etc. 

Let us now illustrate the structure of \eq\nr{eq:1PI} for
the generic model in \eq\nr{eq:generic}. 
Starting from \eq\nr{eq:generic}, writing down indices, 
and employing \eq\nr{eq:dD}, we obtain for 
the right-hand-side of \eq\nr{basicSD},
\ba
\delta_{\phi_i}S \!\!\! & = &  \!\!\! 
-\Delta^{-1}_{ij}\phi_j 
+\fr12 \gamma_{ijk}\Bigl( \phi_j\phi_k + D_{jk} \Bigr) \nn
 \!\!\!  & + &  \!\!\! 
\fr16 \gamma_{ijkl}\Bigl( 
\phi_j\phi_k\phi_l 
+ D_{jk} \phi_l
+ D_{kl} \phi_j
+ D_{lj} \phi_k  
+ D_{jm} D_{kn} D_{lo}
\delta_{\phi_m}
\delta_{\phi_n}
\delta_{\phi_o} S_\rmi{eff}[\phi]
\Bigr). 
\ea
We now take further derivatives according to \eq\nr{eq:1PI}. 
Putting $\phi = 0$ after each differentiation,   
we thus obtain the standard equations
(written in the notation of \eqs\nr{freeprop}--\nr{1pivertex})
\ba
\SDoneLHS(\DLsc) &=&
\sy{}12\SSDoneA(\DLsc,\TAsc,\TAsc)
\sy+16\SSDoneB(\DLsc,\TAsc,\TAsc,\TLsc)\;, \la{eq:SDone}
\\[0ex] 
\SDtwoLHS(\DLsc) &=& 
-\SPropFreeProp(\DLsc) \;
+\sy{}12\SSDtwoA(\DLsc,\TAsc,\TAsc)
\sy+12\SSDtwoB(\DLsc,\DLsc,\TAsc,\TAsc)
\sy+12\SSDtwoC(\DLsc,\TAsc,\TAsc,\TAsc,\TAsc) 
\sy+16\SSDtwoD(\DLsc,\TAsc,\TAsc,\TLsc) \la{eq:SDtwo}
\\[0ex]&=&
-\SPropFreeProp(\DLsc) \;
+\SPropPiProp(\DLsc)\;, 
\\[0ex] \label{3ptSD}
\SDthreeLHS(\DLsc,\DLsc,\DLsc) &=& 
\SDthree(\DLsc,\DLsc,\DLsc)
\sm{+}\SSDthreeA(\DLsc,\DLsc,\DLsc,\TAsc,\TAsc,\TAsc)
\sy+12\SSDthreeB(\DLsc,\DLsc,\DLsc,\TAsc,\TAsc)
\sy+12\!\( \;\; \SSDthreeC(\DLsc,\DLsc,\DLsc,\TAsc,\TAsc)
\sm{+}\SSDthreeD(\DLsc,\DLsc,\DLsc,\TAsc,\TAsc,\TAsc,\TAsc)
\sm{+}\SSDthreeE(\DLsc,\DLsc,\DLsc,\TAsc,\TAsc,\TAsc,\TAsc,\TAsc) 
+\mbox{cyclic}(2,3) \)
\nn[1ex]&&{}
\sm{+}\SSDthreeF(\DLsc,\DLsc,\DLsc,\TAsc,\TAsc,\TAsc,\TAsc,\TLsc)
\sy+12\SSDthreeG(\DLsc,\DLsc,\DLsc,\TAsc,\TAsc,\TAsc,\TAsc)
\sy+16\SSDthreeH(\DLsc,\DLsc,\DLsc,\TAsc,\TAsc,\TLsc)\;,  \la{SDthree}
\\[1ex] 
\SDfourLHS(\DLsc,\DLsc,\DLsc,\DLsc) &=& 
\SDfour(\DLsc,\DLsc,\DLsc,\DLsc) 
\sm{+}\!\( \;\; \SSDfourA(\DLsc,\DLsc,\DLsc,\DLsc,\TAsc,\TAsc,\TAsc,\TAsc)
\sm{+}\SSDfourB(\DLsc,\DLsc,\DLsc,\DLsc,\TAsc,\TAsc,\TAsc) 
\sm{+}\SSDfourD(\DLsc,\DLsc,\DLsc,\DLsc,\TAsc,\TAsc,\TAsc) 
\sy+12\SSDfourE(\DLsc,\DLsc,\DLsc,\DLsc,\TAsc,\TAsc) +\mbox{cyclic}(2,3,4) \)
\sy+12\SSDfourC(\DLsc,\DLsc,\DLsc,\DLsc,\TAsc,\TAsc) 
\nn[1ex]&&{}
+\lb\mbox{2-loop terms}\rb \;, \la{SDfour} \la{eq:SDfive}
\ea
where ``cyclic$(n_1,n_2,...)$'' denotes cyclic permutations of
the legs numbered. 
We have not written down the 2-loop terms in \eq\nr{eq:SDfive}, 
since they are not needed in our explicit 4-loop demonstration below.
Likewise, all higher point 1PI functions $\Gamma_n^\rmi{1PI}$, 
$n\ge 5$, start with 1-loop graphs in the model of \eq\nr{eq:generic}, 
and will again not contribute at this order; 
they will for $\Phi_5$, as well as in the model of~\se\ref{se:lattice}.

Let us stress that in a local theory 
the manipulations needed in~\eq\nr{eq:1PI}
can essentially be made using regular derivatives, 
and can thus easily be implemented algebraically. 
Introducing furthermore $\hbar$ as a loop counting parameter~\cite{CW}, 
allows for an iterative solution of the corresponding SD equations.

\subsection{Vacuum diagrams}

The SD formalism above provides equations relating $n$-point
Green's functions. To incorporate vacuum diagrams, one can use
another simple trick: scaling. Noting that, e.g., $Z[J]$
is a functional of all interaction parameters present in the 
action, $Z[J,\gamma_{ij},\gamma_{ijk},\dots]$, one can derive 
hosts of relations by varying any of these parameters.

A most useful example is to rescale the entire action as
$S[\varphi]\rightarrow \fr1\hbar\, S[\varphi]$, and then vary $\hbar$:
\ba
-\hbar\6_\hbar\,\ln Z[J] &=& \Av{\fr1\hbar\,S[\varphi]} 
\;\;=\;\; \fr1{Z[J]}\, \fr1\hbar\,S[\d_J]\,Z[J] \;.
\ea
Rewriting this in the ``connected'' language (recall $W=\ln Z$),
\ba
-\,\hbar\6_\hbar\,W[J] &=& 
\fr1\hbar\,S[W'[J]+\d_J] \;,
\ea  
allows to finally go over to 1PI functions 
($\6_\hbar W = \6_\hbar S_\rmi{eff} + S'_\rmi{eff} \6_\hbar \phi + J 
\6_\hbar \phi = \6_\hbar S_\rmi{eff}$,
$W'=\phi$ and $\d_J=W''\d_\phi=D[\phi]\delta_\phi$):
\ba \label{preVacSD}
-\,\hbar\6_\hbar\,S_{\rm eff}[\phi] &=& \Av{\fr1\hbar\,S[\varphi]} 
\;\;=\;\; \fr1\hbar\,S[\phi+D[\phi]\d_\phi] \;.
\ea
The free energy $F = - S_{\rm eff}[0]$ can now be 
obtained by setting $\phi=0$ and integrating over $\hbar$.

Noting again that after a rescaling of the integration variables, an expansion
in $\hbar$ is equivalent to the loop expansion~\cite{CW}, one
can integrate the left-hand-side of~\eq\nr{preVacSD} by 
$\int_\hbar (1/\hbar) [...]$, but on the right-hand-side
one integrates over the loop number. Writing
\ba
-S_{\rm eff}[0] &=& F \;\;=\;\; F_0+F_{\rm int}
\;\;=\;\; F_0+\sum_{n\ge 2}F^{\rm int}_{n} \;,
\ea
where $n$ counts the number of loops, it follows that
\ba \label{vacSD}
F^{\rm int}_{n} \;\;=\;\; \fr1{n-1}
\lb \ld \vphantom{I^b} S[\phi+D[\phi]\d_\phi] \right|_{\phi=0} 
\rb_{n}, \quad n\ge 2\;.
\ea
Illustrating \eq\nr{vacSD} for our generic
theory in \eq\nr{eq:generic}, we get 
\ba \label{vacSDillu}
F^{\rm int}_{n} \!\!&=&\!\!\! \fr1{n-1} \!\! \lb \!
\sy-12 \SSDvacO(\TAsc,{})
\sy+16 \SToptVSblobsh(\TAsc,\TAsc,\TLsc)  
\sy+18 \SToptVE(\TAsc,\TAsc)
\sy+18 \SToprVVblobbsh(\TAsc,\TAsc,\TLsc,\TLsc,\TLsc)
\sy+1{24} \SToprVBblobsh(\TAsc,\TAsc,\TAsc,\TAsc) \!\!
\rb_{\! n}\!\! , 
\ea
where we again use the notation of \eqs\nr{freeprop}--\nr{1pivertex}.

In principle the whole loop expansion can 
now be generated from \eq\nr{vacSDillu}, 
using~\eqs\nr{eq:SDtwo}--\nr{eq:SDfive}. 
The $n$-loop vacuum diagrams are
expressed in terms of 1PI $n$-point functions, which in turn are governed
by a set of SD equations. Looking closer at it though, it is 
somewhat of a mess:
one has to expand full propagators in terms of free ones and the $\Pi$'s, use
SD equations to iterate loops for $\Pi$'s, 
which brings back full propagators, etc. Fortunately, 
none of this is necessary for \eq\nr{graphic}, as we now explain.


\section{Generating the irreducible skeletons $\Phi[\Delta]$}
\la{se:combine}

The key observation 
for combining Schwinger-Dyson equations and
the skeleton notation in a useful way
is that we need to extract from~\eq\nr{vacSDillu}
only a specific part, $\Phi[\Delta]$: we already know, 
by~\eq\nr{graphic}, what all the rest combines into. But then, 
full propagators can be replaced by free propagators in all but the 
first term in~\eq\nr{vacSDillu}! Indeed, any self-energy insertion within 
one of the other graphs leads to a two-particle-reducible (2PR) diagram. 
For the same reason, the 1PI vertices in~\eq\nr{vacSDillu}
can be iterated by using
the SD equations of the form in~\eqs\nr{SDthree},\nr{eq:SDfive}, 
but with free propagators! 
More precisely, 
it goes as follows. 

To generate the {\em irreducible} skeletons
$\Phi[\Delta]$ from~\eq\nr{vacSDillu}, it is sufficient to expand 
the first term as
\ba
\SSDvacO(\TAsc,{}) &=& \SSDvacO(\Asc,{}) + \SSDvacOpi(\Asc) 
+\lb\mbox{2PR}\rb 
\nn[1ex]&=&{}
\Tr\, 1 
\sy+12 \SToptVSblobsh(\Asc,\Asc,\Lsc)  
\sy+12 \SToptVE(\Asc,\Asc)
\sy+12 \SToprVVblobbsh(\Asc,\Asc,\Lsc,\Lsc,\Lsc)
\sy+16 \SToprVBblobsh(\Asc,\Asc,\Asc,\Asc) + \{\mbox{2PR}\},
\ea
where in the second step \eq\nr{eq:SDtwo} was used.
Taking into account the minus sign in the relation of $F$ and
$\Phi[\Delta]$, cf.\ \eq\nr{graphic}, and writing
again the loop expansion as $\Phi=\sum_{n\ge 2}\Phi_n$,  
one finally obtains a closed exact equation:
\ba \label{SDvac} 
\Phi_n[\Delta] = \fr1{n-1} \lb 
\sy{}1{12} \SToptVSblobsh(\Asc,\Asc,\Lsc)  
\sy+18 \SToptVE(\Asc,\Asc)
\sy+18 \SToprVVblobbsh(\Asc,\Asc,\Lsc,\Lsc,\Lsc)
\sy+1{24} \SToprVBblobsh(\Asc,\Asc,\Asc,\Asc)
\rb_{n}, \quad n\ge 2\;.
\ea

\eq\nr{SDvac} is our main result. It generates 
all skeletons of all orders in the theory of \eq\nr{eq:generic}, 
once \eqs\nr{SDthree},
\nr{eq:SDfive} are used (with free propagators).
The skeletons, in turn, generate self-energies
via \eq\nr{eq:irrdef} and the analogues of \eqs\nr{red1}--\nr{red3}. 
Inserted finally into~\eq\nr{graphic}, we obtain 
the free energy $F$.
 
\subsection{Vacuum skeletons up to 5-loop level}

The procedure of working out \eq\nr{SDvac}
is simple and mechanical and can, at least up to 4-loop
level, even be carried out by hand, as we shall demonstrate. 
The only complication arising is the identification of 
equivalent topologies: the same graph can be written in very many 
different ways. In order to deal with this situation, 
it appears easiest to assign an algebraic notation for the 
different topologies, rather than a mere graphical one. 
For example, one can count the numbers of 3-point and 4-point vertices
appearing in the graph, and within those equivalence classes, 
one can use a matrix notation for how the vertices are connected.
The significant entries of the matrix can be ordered
to a single number, and by doing the same for all possible
orderings of the vertices, a unique representative
(say, the smallest of such numbers) can be assigned to each topology.    
For an explicit implementation of this kind 
of a procedure, see the 2nd paper in~\cite{Klmany}.

Let us now explicitly work out 
the diagram classes in~\eq\nr{SDvac} up to 4-loop level. For the 
first one, inserting~\eq\nr{SDthree} gives either
a 2-loop graph, or 3-loop graphs to be iterated further on, 
or directly 4-loop graphs:
\ba
\left. \SToptVSblobsh(\Asc,\Asc,\Lsc) \right|_4 \!\! &=& \!\!
\SToptVS(\Asc,\Asc,\Lsc) 
\!\sm{+}\! \lk \SToprVMa(\Asc,\Asc,\Asc,\Lsc,\Lsc,\Lsc)
\sy+12  \SToprVVb(\Asc,\Asc,\Lsc,\Lsc,\Lsc)
\!\sm{+}\!  \SToprVVblobshTxt(\Asc,\Asc,\Lsc,\Lsc,\Lsc,{}) \rk_4
\!\sm{+}\!  \STopfVB(\Asc,\Asc,\Asc,\Asc,\Asc,\Asc,\Lsc)
\!\sm{+}\!  \STopfVV(\Asc,\Asc,\Asc,\Lsc,\Lsc,\Lsc,\Lsc,\Lsc)
\!\sm{+}\!  \STopfVW(\Asc,\Asc,\Asc,\Asc,\Lsc,\Lsc,\Lsc,\Lsc)
\sy+12  \STopfVN(\Asc,\Asc,\Lsc,\Lsc,\Lsc,\Lsc,\Lsc) \,.
\ea
Here the further iterations give
\ba
\left. \SToprVMa(\Asc,\Asc,\Asc,\Lsc,\Lsc,\Lsc) \right|_4 
&=& 
\SToprVM(\Asc,\Asc,\Asc,\Lsc,\Lsc,\Lsc)
\sm{+2} 
\left. \SToprVMb(\Asc,\Asc,\Asc,\Lsc,\Lsc,\Lsc,\mbox{})\right|_4 
= \SToprVM(\Asc,\Asc,\Asc,\Lsc,\Lsc,\Lsc)
\sm{+2} \STopfVH(\Asc,\Asc,\Lsc,\Lsc,\Lsc,\Lsc,\Lsc,\Lsc,\Lsc)
\sm{+3} \STopfVV(\Asc,\Asc,\Asc,\Lsc,\Lsc,\Lsc,\Lsc,\Lsc) \;, \\
\sy{}12 \left. \SToprVVb(\Asc,\Asc,\Lsc,\Lsc,\Lsc) \right|_4
& = &
\sy{}12  \SToprVV(\Asc,\Asc,\Lsc,\Lsc,\Lsc)
\sm{+} \STopfVH(\Asc,\Asc,\Lsc,\Lsc,\Lsc,\Lsc,\Lsc,\Lsc,\Lsc)
\sy+12  \STopfVX(\Asc,\Asc,\Lsc,\Lsc,\Lsc,\Lsc,\Lsc,\Lsc,\Lsc) 
\sm{+2}\! \STopfVW(\Asc,\Asc,\Asc,\Asc,\Lsc,\Lsc,\Lsc,\Lsc)
\sm{+}  \STopfVV(\Asc,\Asc,\Asc,\Lsc,\Lsc,\Lsc,\Lsc,\Lsc)
\sy+12  \STopfVB(\Asc,\Asc,\Asc,\Asc,\Asc,\Asc,\Lsc)
\sy+14  \STopfVU(\Asc,\Asc,\Asc,\Asc,\Lsc,\Lsc,\Lsc) \;, \\
\left. \SToprVVblobshTxt(\Asc,\Asc,\Lsc,\Lsc,\Lsc,{}) \right|_4 
& = & 
\SToprVV(\Asc,\Asc,\Lsc,\Lsc,\Lsc)
\sm{+}  \STopfVV(\Asc,\Asc,\Asc,\Lsc,\Lsc,\Lsc,\Lsc,\Lsc)
\sy+12  \STopfVU(\Asc,\Asc,\Asc,\Asc,\Lsc,\Lsc,\Lsc)
\sm{+}  \STopfVN(\Asc,\Asc,\Lsc,\Lsc,\Lsc,\Lsc,\Lsc) \;.
\ea
We have dropped 5-point functions each time they appear,
since in the model of \eq\nr{eq:generic}, they start 
with a 1-loop term, so that diagrams containing them 
generate higher loop orders. 

The 2nd class in~\eq\nr{SDvac} only contributes to $\Phi_2[\Delta]$, 
and is trivial. 
For the 3rd class in~\eq\nr{SDvac}, 
\ba
\left. \SToprVVblobbsh(\Asc,\Asc,\Lsc,\Lsc,\Lsc) \right|_4 &=&
\SToprVV(\Asc,\Asc,\Lsc,\Lsc,\Lsc) 
\sm{+2}\left. \SToprVVblobshTxt(\Asc,\Asc,\Lsc,\Lsc,\Lsc,\mbox{})\right|_4
=\SToprVV(\Asc,\Asc,\Lsc,\Lsc,\Lsc) 
\sm{+2} \STopfVV(\Asc,\Asc,\Asc,\Lsc,\Lsc,\Lsc,\Lsc,\Lsc)
\sm{+}  \STopfVU(\Asc,\Asc,\Asc,\Asc,\Lsc,\Lsc,\Lsc)
\sm{+2} \STopfVN(\Asc,\Asc,\Lsc,\Lsc,\Lsc,\Lsc,\Lsc) \;.
\ea
For the 4th class, we only need the 1-loop
terms in \eq\nr{SDfour},
\ba
\left. \SToprVBblobsh(\Asc,\Asc,\Asc,\Asc) \right|_4 &=& 
\SToprVB(\Asc,\Asc,\Asc,\Asc)
\sm{+3} \STopfVW(\Asc,\Asc,\Asc,\Asc,\Lsc,\Lsc,\Lsc,\Lsc)
\sm{+6} \STopfVB(\Asc,\Asc,\Asc,\Asc,\Asc,\Asc,\Lsc)
\sy+32  \STopfVT(\Asc,\Asc,\Asc,\Lsc,\Lsc,\Lsc) \;.
\ea 
Collecting finally these different contributions together
with coefficients according to \eq\nr{SDvac}, we get
\ba
\Phi_2 \!\!\!\! &=& \!\!\!  
\sy{}1{12} \SToptVS(\Asc,\Asc,\Lsc)
\sy+18 \SToptVE(\Asc,\Asc)\;, \la{eq:Phi2}
\\[1ex] 
\Phi_3 \!\!\!\! &=& \!\!\! 
\sy{}1{24} \SToprVM(\Asc,\Asc,\Asc,\Lsc,\Lsc,\Lsc)
\sy+18 \SToprVV(\Asc,\Asc,\Lsc,\Lsc,\Lsc)
\sy+1{48} \SToprVB(\Asc,\Asc,\Asc,\Asc)\;, \la{eq:Phi3}
\\[1ex] 
\Phi_4 \!\!\!\! &=& \!\!\! 
\sy{}1{72}\!  \STopfVX(\Asc,\Asc,\Lsc,\Lsc,\Lsc,\Lsc,\Lsc,\Lsc,\Lsc) 
\!\sy+1{12}\! \STopfVH(\Asc,\Asc,\Lsc,\Lsc,\Lsc,\Lsc,\Lsc,\Lsc,\Lsc)
\!\sy+18\!   \STopfVW(\Asc,\Asc,\Asc,\Asc,\Lsc,\Lsc,\Lsc,\Lsc)
\!\sy+14\!   \STopfVV(\Asc,\Asc,\Asc,\Lsc,\Lsc,\Lsc,\Lsc,\Lsc) 
\!\sy+18\!   \STopfVB(\Asc,\Asc,\Asc,\Asc,\Asc,\Asc,\Lsc)
\!\sy+18\!   \STopfVN(\Asc,\Asc,\Lsc,\Lsc,\Lsc,\Lsc,\Lsc)
\!\sy+1{16}\! \STopfVU(\Asc,\Asc,\Asc,\Asc,\Lsc,\Lsc,\Lsc)
\!\sy+1{48}\! \STopfVT(\Asc,\Asc,\Asc,\Lsc,\Lsc,\Lsc)\;. \la{eq:Phi4}
\ea

Proceeding to higher loop orders, an automatised treatment proves
essential, for the reasons outlined above. Implementing our generic
formulae as well as an ordering algorithm
separating topologies in {\tt FORM}~\cite{jamv}, we obtain
in a straightforward way the complete set of
5-loop skeletons,
\ba
\Phi_5 \!\!\!\! &=& \!\!\! 
 \sy{}1{4}\!  \STopEVa(\Asc,\Lsc)
\!\sy+1{48}\! \STopEVb(\Asc,\Lsc) 
\!\sy+1{16}\! \STopEVc(\Asc,\Lsc) 
\!\sy+1{12}\! \STopEVd(\Asc,\Lsc)
\!\sy+1{4}\! \STopEVe(\Asc,\Lsc)
\!\sy+1{2}\! \STopEVf(\Asc,\Lsc)
\!\sy+1{2}\! \STopEVg(\Asc,\Lsc)
\nn[1ex]&&{}
\!\sy+1{8}\! \STopEVh(\Asc,\Lsc)
\!\sy+1{4}\! \STopEVi(\Asc,\Lsc)
\!\sy+1{4}\! \STopEVj(\Asc,\Lsc)
\!\sy+1{8}\! \STopEVk(\Asc,\Lsc)
\!\sy+1{8}\! \STopEVl(\Asc,\Lsc)
\!\sy+1{4}\! \STopEVm(\Asc,\Lsc)
\!\sy+1{4}\! \STopEVn(\Asc,\Lsc)
\nn[1ex]&&{}
\!\sy+1{8}\! \STopEVo(\Asc,\Lsc)
\!\sy+1{2}\! \STopEVp(\Asc,\Lsc)
\!\sy+1{8}\! \STopEVq(\Asc,\Lsc)
\!\sy+1{4}\! \STopEVr(\Asc,\Lsc)
\!\sy+1{16}\! \STopEVs(\Asc,\Lsc)
\!\sy+1{8}\! \STopEVt(\Asc,\Lsc)
\!\sy+1{4}\! \STopEVu(\Asc,\Lsc)
\nn[1ex]&&{}
\!\sy+1{2}\! \STopEVv(\Asc,\Lsc)
\!\sy+1{16}\! \STopEVw(\Asc,\Lsc)
\!\sy+1{12}\! \STopEVx(\Asc,\Lsc)
\!\sy+1{16}\! \STopEVy(\Asc,\Lsc)
\!\sy+1{32}\! \STopEVz(\Asc,\Lsc)
\!\sy+1{16}\! \STopEVaa(\Asc,\Lsc)
\!\sy+1{8}\! \STopEVab(\Asc,\Lsc)
\nn[1ex]&&{}
\!\sy+1{4}\! \STopEVac(\Asc,\Lsc)
\!\sy+1{8}\! \STopEVad(\Asc,\Lsc)
\!\sy+1{4}\! \STopEVae(\Asc,\Lsc)
\!\sy+1{8}\! \STopEVaf(\Asc,\Lsc)
\!\sy+1{12}\! \STopEVag(\Asc,\Lsc)
\!\sy+1{128}\! \STopEVah(\Asc,\Lsc)
\!\sy+1{32}\! \STopEVai(\Asc,\Lsc)
\;. \la{eq:Phi5}
\ea
Note once more that these skeletons are all that is needed for 
generating the loop expansion for the full free energy, as discussed 
above. 

\subsection{Self-energies up to 2-loop level} 
\la{se:genericSE}

Now that we have $\Phi_n[\Delta]$ in~\eqs\nr{eq:Phi2}--\nr{eq:Phi5},  
irreducible as well as reducible self-energies can easily be 
obtained with \eqs\nr{eq:irrdef}, \nr{red1}--\nr{red3}, etc. 
For bosonic particles, for instance ($c_\rmi{i}=\fr12$), we get
\ba \la{genericSE}
\Pi_1^{\rm irr} = \SPropCircProp(\DLsc,1) &=& 
\sy{}12 \STopoSB(\DLsc,\Asc,\Asc) 
\sy+12  \STopoST(\DLsc,\Asc) \;, 
\\[0ex]
\Pi_2^{\rm irr} = \ToptSi(\DLsc) &=& 
\sy{}12 \SToptSM(\DLsc,\Asc,\Asc,\Asc,\Asc,\Lsc) 
\sy+12  \SToptSAl(\DLsc,\Asc,\Asc,\Asc,\Asc) 
\sy+12  \SToptSAr(\DLsc,\Asc,\Asc,\Asc,\Asc) 
\sy+14  \SToptSE(\DLsc,\Asc,\Asc,\Asc,\Asc) 
\sy+16  \SToptSS(\DLsc,\Asc,\Asc,\Lsc) \;,
\\[0ex]
\Pi_2^{\rm red(1)} = \ToptSr(\DLsc) &=& 
\sm{1} \SToptSBB(\DLsc,\Asc,\Asc) 
\sy+12 \SToptSTB(\DLsc,\Asc) \;, \la{pi2r1}
\ea
etc.
Note that the outcome of the derivative in \eq\nr{eq:irrdef}
must be symmetric in all (bosonic) indices. 
The 3- and 4-loop self-energies could be derived from $\Phi_4$
and $\Phi_5$, respectively,
but we choose not to give them here, since they are not needed for the
set of 4-loop vacuum diagrams that we will 
display explicitly in \se\ref{se:continuum}.

With \eqs\nr{genericSE}--\nr{pi2r1}, the ring diagrams 
in \eq\nr{graphic} are readily written down.


\section{Generic model on the lattice}
\la{se:lattice}

So far we have considered the generic model in \eq\nr{eq:generic}. 
However, in a lattice regularization of gauge theories, higher 
vertices appear as well, without spoiling renormalizability. 
At the generic level, it is straightforward to add  
such couplings to the theory in \eq\nr{eq:generic}.
We can include, e.g., 
terms up to 
$\sim (1/8!)\gamma_{ijklmnop}
\varphi_i \varphi_j \varphi_k \varphi_l 
\varphi_m \varphi_n \varphi_o \varphi_p$, as would arise
in lattice perturbation theory for SU($N$) gauge theories, if one 
keeps terms contributing to 4-loop vacuum graphs. Such computations 
would be needed when one converts results of three-dimensional 
numerical Monte Carlo studies from lattice to continuum
regularization~\cite{framework}. 

In this case, everything goes as before, except for the 
appearance of extra vertices in the SD equations, 
as well as in~\eq\nr{SDvac}. 
We shall here simply spell out the final results, 
without rewriting explicitly the modified SD equations. 
We obtain the following additional skeletons,
\def\lat{\left.\vphantom{1^1_1}\right|_{\rm lat}}
\ba
\Phi_3\lat &=& 
\sy{}1{12} \STopLA(\Asc,\Lsc) 
\sy+1{48}  \STopLB(\Asc,\Lsc) \;, \\[1ex]
\Phi_4\lat &=& 
\sy{}1{8}  \STopLa(\Asc,\Lsc) 
\sy+1{12}  \STopLb(\Asc,\Lsc) 
\sy+1{240} \STopLc(\Asc,\Lsc) 
\sy+1{12}  \STopLd 
\sy+1{8}   \STopLe 
\sy+1{16}  \STopLf \nn[1ex]&&{}
\sy+1{48}  \STopLg 
\sy+1{72}  \STopLh 
\sy+1{48}  \STopLi 
\sy+1{48}  \STopLj 
\sy+1{384} \STopLk \;,
\ea
as well as the additional irreducible self-energy, 
\ba
\Pi_2^{\rm irr}\lat = \ToptSi(\DLsc)\lat &=&
\sy+14 \spicb{\DLsc(0,15)(7.5,15) \Asc(22.5,15)(15,0,360)
\DLsc(37.5,15)(45,15) \Asc(15,15)(7.5,0,360)}
\sy+14 \spicb{\DLsc(0,15)(7.5,15) \Asc(22.5,15)(15,0,360)
\DLsc(37.5,15)(45,15) \Asc(30,15)(7.5,0,360)}
\sy+16 \spicb{\DLsc(0,0)(45,0) \Asc(22.5,15)(15,0,360) \Lsc(22.5,0)(22.5,30)} 
\sy+18 \spicb{\DLsc(0,15)(45,15) \Asc(22.5,7)(8,0,360) \Asc(22.5,23)(8,0,360)}
\;,
\ea
where we again assumed $c_\rmi{i} = \fr12$.

\section{Applications: QCD, QED, SQED, electroweak theory}
\la{se:continuum}

As an application of the generic formulae derived above, 
we consider in this section
SU($N$) gauge theory with fermions and a scalar field.
This class includes QCD and QED (where graphs containing
scalar propagators and, in the latter case, gauge field self-interactions, 
are to be dropped out), as well as 
the electroweak theory and scalar electrodynamics (SQED). 
For brevity, we display here only the vertices appearing
in the symmetric phases of the latter theories. 
We mostly use the language of QCD, referring to the
gauge fields as gluons, etc.

The Lagrangian is specified by giving Feynman rules 
for the free propagators and free vertices,
\def\BareProp(#1){\pic{#1(0,15)(30,15)}}
\def\SBareProp(#1){\spic{#1(0,15)(30,15)}}
\ba
\SBareProp(\Lgl) \;,\;
\SBareProp(\Lgh) \;,\;
\SBareProp(\Lsc) \;,\;
\SSDthree(\Lgl,\Lgl,\Lgl) \;,\; 
\SSDthree(\Lgl,\Lgh,\Lagh) \;,\; 
\SSDthree(\Lgl,\Lsc,\Lsc) \;,\; 
\SSDfour(\Lgl,\Lgl,\Lgl,\Lgl) \;,\;
\SSDfour(\Lgl,\Lgl,\Lsc,\Lsc) \;,\;
\SSDfour(\Lsc,\Lsc,\Lsc,\Lsc) \;,  \la{Frules}
\ea
where gluons/scalars are denoted by wavy/straight lines.
Both quarks and ghosts are denoted here by dotted lines;
the Feynman rules for them are different, but the 
symmetry factors agree --- the only exception being diagrams
with more than one closed fermion loop, in which case both
ghosts and quarks can appear in the same diagram
simultaneously, reducing the
symmetry by an obvious factor.  

We do not here write down counterterms explicitly. 
Coupling constant counterterms can be viewed
as a part of the cubic and quartic couplings, while wave 
function and mass counterterms can be treated as a part
of the {\em irreducible} self-energies $\Pi_n^\rmi{irr}$, making their
appearance only in ring diagrams
according to~\eq\nr{graphic}. 

Let us first note that once we write down the
summation over the field 
content explicitly in~\eq\nr{eq:generic}, 
the ``natural'' symmetry factors
in front of the vertices change. For instance, writing
the 4-point vertex in the case of two sets of fields, 
$\{ \varphi_i \} \to \{ A_i \} + \{ B_\alpha \}$, and 
using the symmetry of $\gamma_{ijkl}$, one gets
\be
\frac{1}{4!}\gamma_{ijkl} \varphi_i \varphi_j \varphi_k \varphi_l = 
\frac{1}{4!}\gamma_{ijkl} A_i A_j A_k A_l
+ \frac{1}{3!}\gamma_{ijk\alpha} A_i A_j A_k B_\alpha
+ \frac{1}{(2!)^2}\gamma_{ij\alpha\beta} A_i A_j B_\alpha B_\beta + ... \;.
\ee
Similarly, writing the 3-point vertex for three different fields, 
$\{ \varphi_i \} \to \{ A_i \} + \{B_\alpha\} + \{C_M \}$, one finds
\be
\frac{1}{3!}\gamma_{ijk} \varphi_i \varphi_j \varphi_k = 
\frac{1}{3!}\gamma_{ijk} A_i A_j A_k + 
\frac{1}{2!}\gamma_{ij\alpha} A_i A_i B_\alpha + 
\gamma_{i \alpha M} A_i B_\alpha C_M + ... \;.
\ee
With these conventions, 
each tree-level vertex in the graphical notation corresponds 
just to $\gamma_{ijkl}, \gamma_{ijk\alpha}$, etc, without 
any symmetry factors there: all of them are shown explicitly. 

The only thing remaining is to write the 
summation over particle species explicitly also 
in the propagators of~\eqs\nr{eq:Phi2}--\nr{eq:Phi4}, 
\be
\SBareProp(\Lsc)  \equiv 
\SBareProp(\Lgl)  \;+\;
\SBareProp(\Lgh)  \;+\;
\SBareProp(\Lagh) \;+\;
\SBareProp(\Lsc)  \;.
\ee
Only the vertices allowed by the Feynman rules are kept
after this substitution. 
This generates all the graphs, with the correct symmetry factors.

\subsection{Vacuum skeletons up to 4-loop level}

The procedure outlined above can easily be carried out explicitly,  
and up to 4-loop level even by hand. The main complication is again 
the identification of various equivalent topologies, and for 
this a suitable algebraic notation may be more useful than 
a graphical one.  As a result, 
for the field content in~\eq\nr{Frules}, we finally obtain
\ba
\Phi_2 &=&  
\sy{}18 \SToptVE(\Agl,\Agl)
\sy+1{12} \SToptVS(\Agl,\Agl,\Lgl)  
\sy-12 \SToptVS(\Agh,\Agh,\Lgl)
\sy+14 \SToptVE(\Agl,\Asc)
\sy+14 \SToptVS(\Asc,\Asc,\Lgl)
\plus\sy{}18 \SToptVE(\Asc,\Asc) \;,
\\[1ex] 
\Phi_3 &=& 
\sy{}1{24} \SToprVM(\Agl,\Agl,\Agl,\Lgl,\Lgl,\Lgl)
\sy-13 \SToprVM(\Agh,\Agh,\Agh,\Lgl,\Lgl,\Lgl)
\sy-14 \SToprVM(\Agh,\Agl,\Agh,\Lhh,\Lhh,\Lgl)
\sy+18 \SToprVV(\Agl,\Agl,\Lgl,\Lgl,\Lgl)
\sy+1{48} \SToprVB(\Agl,\Agl,\Agl,\Agl)
\sy+16 \SToprVM(\Asc,\Asc,\Asc,\Lgl,\Lgl,\Lgl)
\sy+18\SToprVM(\Asc,\Agl,\Asc,\Lsc,\Lsc,\Lgl) 
\nn[1ex]&&{}
\sy+12\SToprVV(\Asc,\Asc,\Lsc,\Lgl,\Lgl)  
\sy+14\SToprVV(\Agl,\Asc,\Lgl,\Lgl,\Lsc)  
\sy+18\SToprVB(\Asc,\Asc,\Agl,\Agl) 
\plus\sy{}18\SToprVV(\Asc,\Asc,\Lgl,\Lsc,\Lsc) 
\plus\sy{}1{48}\SToprVB(\Asc,\Asc,\Asc,\Asc) \;,
\\[1ex] 
\Phi_4 &=& \!\!
\sy{}1{72} \STopfVX(\Agl,\Agl,\Lgl,\Lgl,\Lgl,\Lgl,\Lgl,\Lgl,\Lgl) 
\sy-14     \STopfVX(\Agh,\Agh,\Lhh,\Lhh,\Lgl,\Lgl,\Lgl,\Lgl,\Lgl)
\sy-16     \STopfVX(\Agh,\Agh,\Lgl,\Lhh,\Lgl,\Lhh,\Lhh,\Lgl,\Lhh)
\sy+1{12}\STopfVH(\Agl,\Agl,\Lgl,\Lgl,\Lgl,\Lgl,\Lgl,\Lgl,\Lgl)
\sy-12   \STopfVH(\Agh,\Agh,\Lhh,\Lhh,\Lgl,\Lgl,\Lgl,\Lgl,\Lgl)
\sy-12   \STopfVH(\Agh,\Agh,\Lgl,\Lhh,\Lgl,\Lhh,\Lhh,\Lgl,\Lhh)
\nn[1ex]&&{} \!\!
\sm{-1}  \STopfVH(\Agh,\Agl,\Lhh,\Lhh,\Lgl,\Lgl,\Lhh,\Lhh,\Lgl)
\sy-13   \STopfVH(\Agh,\Agl,\Lgl,\Lgl,\Lhh,\Lhh,\Lgl,\Lgl,\Lgl)
\sy+16   \STopfVH(\Agh,\Agh,\Lgl,\Lgl,\Lhh,\Lhh,\Lhh,\Lhh,\Lgl)
\sy+16   \STopfVH(\Agh,\Aagh,\Lgl,\Lgl,\Lhh,\Lhh,\Lhh,\Lhh,\Lgl)
\sy+18   \STopfVW(\Agl,\Agl,\Agl,\Agl,\Lgl,\Lgl,\Lgl,\Lgl)
\sy-14   \STopfVW(\Agh,\Agh,\Agh,\Agh,\Lgl,\Lgl,\Lgl,\Lgl)
\nn[1ex]&&{} \!\!
\sy+14   \STopfVV(\Agl,\Agl,\Agl,\Lgl,\Lgl,\Lgl,\Lgl,\Lgl)
\sy-12   \STopfVV(\Agl,\Agl,\Agh,\Lgl,\Lhh,\Lhh,\Lgl,\Lgl) 
\sy+18   \STopfVB(\Agl,\Agl,\Agl,\Agl,\Agl,\Agl,\Lgl)
\sy+18   \STopfVN(\Agl,\Agl,\Lgl,\Lgl,\Lgl,\Lgl,\Lgl)
\sy+1{16}\STopfVU(\Agl,\Agl,\Agl,\Agl,\Lgl,\Lgl,\Lgl)
\sy+1{48}\STopfVT(\Agl,\Agl,\Agl,\Lgl,\Lgl,\Lgl)
\nn[1ex]&&{} \!\!
%
%
\sy+18 \STopfVX(\Asc,\Asc,\Lsc,\Lsc,\Lgl,\Lgl,\Lgl,\Lgl,\Lgl)
\sy+1{12} \STopfVX(\Asc,\Asc,\Lgl,\Lsc,\Lgl,\Lsc,\Lsc,\Lgl,\Lsc)
\sy-13\STopfVH(\Asc,\Agh,\Lgl,\Lgl,\Lsc,\Lsc,\Lhh,\Lhh,\Lgl)
\sy+14\STopfVH(\Asc,\Asc,\Lsc,\Lsc,\Lgl,\Lgl,\Lgl,\Lgl,\Lgl)
\sy+14\STopfVH(\Asc,\Asc,\Lgl,\Lsc,\Lgl,\Lsc,\Lsc,\Lgl,\Lsc)
\sy+12\STopfVH(\Asc,\Agl,\Lsc,\Lsc,\Lgl,\Lgl,\Lsc,\Lsc,\Lgl)
\nn[1ex]&&{} \!\!
\sy+16\STopfVH(\Asc,\Agl,\Lgl,\Lgl,\Lsc,\Lsc,\Lgl,\Lgl,\Lgl)
\sy+1{12}\STopfVH(\Asc,\Asc,\Lgl,\Lgl,\Lsc,\Lsc,\Lsc,\Lsc,\Lgl) 
\sy+12\STopfVW(\Asc,\Agl,\Agl,\Agl,\Lsc,\Lsc,\Lgl,\Lgl)
\sy+12\STopfVW(\Asc,\Agl,\Agl,\Asc,\Lgl,\Lsc,\Lgl,\Lsc)
\sy+12\STopfVW(\Agl,\Asc,\Asc,\Asc,\Lsc,\Lsc,\Lgl,\Lgl) 
\sy+18\STopfVW(\Asc,\Asc,\Asc,\Asc,\Lgl,\Lgl,\Lgl,\Lgl) 
\sy+14\STopfVV(\Asc,\Agl,\Agl,\Lgl,\Lgl,\Lgl,\Lsc,\Lgl) 
\nn[1ex]&&{} \!\!
\sy+14\STopfVV(\Asc,\Agl,\Asc,\Lgl,\Lsc,\Lsc,\Lsc,\Lgl)
\sy-12\STopfVV(\Asc,\Agl,\Agh,\Lgl,\Lhh,\Lhh,\Lsc,\Lgl) 
\sy+14\STopfVV(\Agl,\Agl,\Asc,\Lgl,\Lsc,\Lsc,\Lgl,\Lgl)
\sy+14\STopfVV(\Agl,\Asc,\Agl,\Lgl,\Lsc,\Lgl,\Lgl,\Lsc)
\sy+14\STopfVV(\Agl,\Asc,\Asc,\Lgl,\Lgl,\Lsc,\Lgl,\Lsc)
\sm{+1}\STopfVV(\Asc,\Asc,\Asc,\Lsc,\Lgl,\Lgl,\Lgl,\Lgl) 
\sm{+1}\STopfVV(\Asc,\Asc,\Agl,\Lsc,\Lsc,\Lsc,\Lgl,\Lgl) 
\nn[1ex]&&{} \!\!
\sy+14\STopfVB(\Agl,\Agl,\Asc,\Agl,\Agl,\Asc,\Lsc)
\sy+18\STopfVB(\Agl,\Agl,\Asc,\Asc,\Asc,\Asc,\Lgl)
\sy+12\STopfVB(\Asc,\Agl,\Asc,\Asc,\Agl,\Agl,\Lgl)
\sy+12\STopfVB(\Asc,\Agl,\Agl,\Asc,\Agl,\Asc,\Lsc)
\sy+18\STopfVB(\Asc,\Asc,\Agl,\Agl,\Agl,\Agl,\Lgl) 
\sy+14\STopfVN(\Asc,\Agl,\Lgl,\Lgl,\Lsc,\Lgl,\Lgl)
\nn[1ex]&&{} \!\!
\sy+18\STopfVN(\Asc,\Asc,\Lgl,\Lgl,\Lsc,\Lsc,\Lgl)
\sy+12\STopfVN(\Agl,\Asc,\Lgl,\Lsc,\Lgl,\Lgl,\Lsc)
\sy+12\STopfVN(\Asc,\Asc,\Lsc,\Lsc,\Lgl,\Lgl,\Lgl) 
\sy+18\STopfVU(\Agl,\Asc,\Agl,\Agl,\Lgl,\Lsc,\Lgl)
\sy+1{16}\STopfVU(\Agl,\Agl,\Agl,\Asc,\Lgl,\Lgl,\Lsc)
\sy+12\STopfVU(\Asc,\Asc,\Asc,\Asc,\Lgl,\Lgl,\Lgl)
\sy+1{16}\STopfVU(\Asc,\Asc,\Agl,\Agl,\Lsc,\Lsc,\Lgl)
\nn[1ex]&&{} \!\!
\sy+1{16}\STopfVT(\Agl,\Asc,\Agl,\Lgl,\Lgl,\Lsc)
\sy+16\STopfVT(\Asc,\Asc,\Asc,\Lgl,\Lgl,\Lgl) 
\plus\sy{}14\STopfVW(\Agl,\Agl,\Asc,\Asc,\Lsc,\Lsc,\Lsc,\Lsc) 
\sy+14\STopfVV(\Asc,\Asc,\Agl,\Lgl,\Lsc,\Lgl,\Lsc,\Lsc)
\sy+14\STopfVV(\Asc,\Asc,\Asc,\Lgl,\Lgl,\Lsc,\Lsc,\Lsc)
\sy+14\STopfVB(\Asc,\Asc,\Asc,\Agl,\Agl,\Asc,\Lsc)
\sy+12\STopfVN(\Asc,\Agl,\Lgl,\Lsc,\Lsc,\Lsc,\Lsc)
\nn[1ex]&&{} \!\!
\sy+18\STopfVU(\Agl,\Asc,\Agl,\Asc,\Lgl,\Lsc,\Lsc)
\sy+1{16}\STopfVT(\Asc,\Agl,\Asc,\Lsc,\Lsc,\Lgl)
\plus\sy{}18\STopfVB(\Asc,\Asc,\Asc,\Asc,\Asc,\Asc,\Lgl) 
\sy+1{16}\STopfVU(\Asc,\Asc,\Agl,\Asc,\Lsc,\Lsc,\Lsc)
\plus\sy{}1{48}\STopfVT(\Asc,\Asc,\Asc,\Lsc,\Lsc,\Lsc) \;.
\ea


\subsection{Self-energies up to 2-loop level}

Using \eqs\nr{eq:irrdef}, \nr{red1}, the skeletons above immediately 
produce the self-energies of the model in \eq\nr{Frules}. 
We obtain 
\ba
%
%
\TopoS(\Lgl) &=&
\sy{}12 \STopoSB(\Lgl,\Agl,\Agl)
\sm{-1} \STopoSB(\Lgl,\Agh,\Agh)
\sy+12 \STopoST(\Lgl,\Agl)
\sy+12 \STopoSB(\Lgl,\Asc,\Asc)
\sy+12 \STopoST(\Lgl,\Asc) \;,
\\[0ex]
\TopoS(\Lgh) &=&
\sm{1} \STopoSB(\Lhh,\Agl,\Agh) \;,
\\[0ex]
\TopoS(\Lsc) &=& 
\sm{1} \STopoSB(\Lsc,\Asc,\Agl) 
\sy+12 \STopoST(\Lsc,\Agl)  
\plus\sy{}12 \STopoST(\Lsc,\Asc) \;,
\\[0ex]
%
%
\ToptSi(\Lgl) &=& 
\sy{}12 \SToptSM(\Lgl,\Agl,\Agl,\Agl,\Agl,\Lgl)
\sm{-1} \SToptSM(\Lgl,\Agh,\Agl,\Agl,\Agh,\Lgh)
\sm{-1} \SToptSM(\Lgl,\Agl,\Agh,\Agh,\Agl,\Lagh)
\sm{-1} \SToptSM(\Lgl,\Agh,\Agh,\Agh,\Agh,\Lgl)
\sy+12 \SToptSAl(\Lgl,\Agl,\Agl,\Agl,\Agl)
\sy+12 \SToptSAr(\Lgl,\Agl,\Agl,\Agl,\Agl) \nn[0ex]&&{}
\sy+14 \SToptSE(\Lgl,\Agl,\Agl,\Agl,\Agl)
\sy+16 \SToptSS(\Lgl,\Agl,\Agl,\Lgl)
\sy+12 \SToptSM(\Lgl,\Asc,\Agl,\Agl,\Asc,\Lsc)
\sy+12 \SToptSM(\Lgl,\Agl,\Asc,\Asc,\Agl,\Lsc)
\sy+12  \SToptSM(\Lgl,\Asc,\Asc,\Asc,\Asc,\Lgl) \nn[0ex] && {}
\sm{+1} \SToptSAl(\Lgl,\Asc,\Asc,\Asc,\Agl)
\sm{+1} \SToptSAr(\Lgl,\Asc,\Asc,\Asc,\Agl)  
\sy+12 \SToptSAl({\SetColor{Black}\Lgl},\Agl,\Asc,\Agl,\Asc)
\sy+12 \SToptSAr({\SetColor{Black}\Lgl},\Asc,\Agl,\Agl,\Asc) \nn[0ex] && {}
\sy+14 \SToptSE({\SetColor{Black}\Lgl},\Asc,\Agl,\Agl,\Asc)
\sy+14 \SToptSE({\SetColor{Black}\Lgl},\Agl,\Asc,\Asc,\Agl)
\sy+12 \SToptSS(\Lgl,\Asc,\Asc,\Lgl) 
\plus\sy{}14  \SToptSE({\SetColor{Black}\Lgl},\Asc,\Asc,\Asc,\Asc) \;,
\\[0ex]
\ToptSi(\Lgh) &=&
\sm{1} \SToptSM(\Lhh,\Agl,\Agl,\Agh,\Agh,\Lgl)
\sm{+1} \SToptSM(\Lhh,\Aagh,\Agl,\Agh,\Agl,\Lagh) \;,
\\[0ex]
\ToptSi(\Lsc) &=&
\sm{1} \SToptSM(\Lsc,\Agl,\Agl,\Asc,\Asc,\Lgl)
\sm{+1} \SToptSM(\Lsc,\Asc,\Agl,\Asc,\Agl,\Lsc)
\sy+12 \SToptSAl({\SetColor{Black}\Lsc},\Agl,\Agl,\Asc,\Agl)
\sy+12 \SToptSAr({\SetColor{Black}\Lsc},\Agl,\Agl,\Asc,\Agl) 
\sm{+1} \SToptSAl(\Lsc,\Asc,\Asc,\Agl,\Agl)
\sm{+1} \SToptSAr(\Lsc,\Asc,\Asc,\Agl,\Agl) \nn[0ex] && {}
\sm{+1} \SToptSE(\Lsc,\Agl,\Agl,\Asc,\Asc)
\sy+12 \SToptSS(\Lsc,\Agl,\Agl,\Lsc) 
\plus\sy{}12 \SToptSAl({\SetColor{Black}\Lsc},\Agl,\Asc,\Asc,\Asc)
\sy+12 \SToptSAr({\SetColor{Black}\Lsc},\Asc,\Agl,\Asc,\Asc) 
\plus\sy{}16 \SToptSS(\Lsc,\Asc,\Asc,\Lsc) \;,
\\[0ex]
%
%
\ToptSr(\Lgl ) &=&
\sm{1} \SToptSBB(\Lgl,\Agl,\Agl)
\sm{-1} \SToptSBB(\Lgl,\Agh,\Agh)
\sm{-1} \SToptSBB(\Lgl,\Aagh,\Aagh)
\sy+12 \SToptSTB(\Lgl,\Agl)
\sm{+1}\SToptSBB(\Lgl,\Asc,\Asc)
\sy+12\SToptSTB(\Lgl,\Asc) \;,
\\[0ex]
\ToptSr(\Lgh) &=&
\sm{1} \SToptSBB(\Lhh,\Agl,\Agh)
\sm{+1} \SToptSBB(\Lhh,\Aagh,\Agl) \;,
\\[0ex]
\ToptSr(\Lsc)&=&
\sm{1}\SToptSBB(\Lsc,\Agl,\Asc)
\sm{+1}\SToptSBB(\Lsc,\Asc,\Agl)
\sy+12\SToptSTB(\Lsc,\Agl)
\sy+12\SToptSTB(\Lsc,\Asc) \;.
\ea


\subsection{Ring diagrams up to 4-loop level}

To be exhaustive up to 4-loop level, let us finally give the set of 
ring diagrams for the model of~\eq\nr{Frules}. 
While there are no ring diagrams up to 2-loop level, from \eq\nr{graphic} 
we get
\ba 
\(-F_{\rm (rings)}\)_3 &=&
\sy{}14  \STTopoVRoo(\Agl)
\sy-12  \STTopoVRoo(\Agh)
\sy+14  \STTopoVRoo(\Asc) \;,
\\[1ex]
\(-F_{\rm (rings)}\)_4 &=&
\sy{}16  \STTopoVRooo(\Agl)
\sy+12 \STTopoVRoi(\Agl)
\sy+14  \STTopoVRor(\Agl)
\sy-13  \STTopoVRooo(\Agh)
\sm{-1} \STTopoVRoi(\Agh)
\sy-12  \STTopoVRor(\Agh)
\nn[1ex]&&{}
\sy+16  \STTopoVRooo(\Asc)
\sy+12  \STTopoVRoi(\Asc) 
\sy+14  \STTopoVRor(\Asc) \;.
\ea
Note the extremely economic structure of the skeleton expansion 
of \eq\nr{graphic}: the few ring diagrams above summarize 22 (276) 
3-loop (4-loop) diagrams.


\section{Discussion}
\la{se:integrals}

In this paper we have described a simple practical procedure
for systematically generating all vacuum diagrams
of a given loop order in a generic field theory. 

We have shown that the sum of vacuum diagrams can be written 
in the form of a modified skeleton expansion, \eq\nr{graphic}. It 
contains 2-particle-irreducible ``skeletons'' with free propagators, 
as well as various self-energy insertions
inside ``ring diagrams''. The self-energies are, 
in turn, determined by the skeletons. Thus, all one 
really needs is the skeletons. 

The 2-particle-irreducible skeletons of a given order are, 
then, generated by~\eq\nr{SDvac}. It contains a number of 
full 3-point and 4-point vertices, which can in turn be 
expanded using specific ``irreducible'' Schwinger-Dyson 
equations (\eqs\nr{SDthree}, \nr{eq:SDfive}, etc), where
full propagators have been replaced with free propagators.
In this way, all vacuum graphs are generated simultaneously, 
with the correct symmetry factors. Finally, the precise particle 
content of the theory one is interested in can be specified as 
discussed in~\se\ref{se:continuum}.
Our method is also directly applicable to theories with
spontaneous symmetry breaking, as only free propagators 
and vertices are modified; tadpole graphs 
are generated by~\eq\nr{eq:tadpoles}.

This iterative procedure is very 
straightforward and can be automatised, 
but up to 4-loop level the computations are 
easily carried out even by hand, as we have demonstrated. 
Thus, we believe that our setup economises 
the generation of the set of high order vacuum diagrams, 
compared with techniques where all the types of graphs have to 
be dealt with on the same footing, without a separation into 
skeletons with free propagators, and ring diagrams.  

Up to this point, 
we have not discussed at all the integrations remaining to be 
carried out, after the diagrams have been generated.
Let us end by pointing out that our setup is beneficial as far as 
their structure is considered, as well, 
in dimensions lower than four~\cite{ma}. 

The point is that low-dimensional field theories 
of the type in~\eq\nr{eq:generic}
are super-renorma\-lizable.
In fact, for $d=2,3$, only the 2-point function suffers from ultraviolet 
divergences, as can be seen by simple power counting. 
Therefore the skeleton graphs which by definition do not have any genuine
2-point functions inside them, do not contain any ultraviolet
divergences in subdiagrams. The ring diagrams, on the other hand, 
do have ultraviolet divergences in subdiagrams.
Note, in particular, that
since $\Pi_n^\rmi{irr}, \Pi_n^{\rmi{red}(m)}$ come with different
symmetry factors in~\eq\nr{graphic}, the counterterms in 
$\Pi_n^\rmi{irr}$ which make the whole $\Pi_n$ finite, do not
in general immediately cancel all the 
ultraviolet subdivergences of the ring diagrams. 

Consequently, various ring diagram classes can 
contribute to the overall divergences of the vacuum 
graphs with potentially infrared sensitive coefficients, 
coming from the other parts of the final integration, 
while skeleton diagrams can not.  
Fortunately, the ring diagram integrations are simpler than those 
in the skeleton graphs, and this problem can thus be dealt with 
in a tractable setting~\cite{inpre}.   


\section*{Acknowledgements}

We thank M.~Achhammer, U.~Heinz, S.~Leupold and H.~Schulz 
for useful discussions
and correspondence. This work was partly supported by the TMR network 
{\em Finite Temperature Phase Transitions in Particle Physics}, 
EU contract no.\ FMRX-CT97-0122, by the RTN network {\em Supersymmetry 
and the Early Universe}, EU contract no.\ HPRN-CT-2000-00152, 
and by the Academy of Finland, project no.\ 163065. 




\begin{thebibliography}{99}

\bibitem{qed}
V.W.~Hughes and T.~Kinoshita,
Rev.\ Mod.\ Phys.\  {71} (1999) S133;
%
A.~Czarnecki and W.J.~Marciano,
Nucl.\ Phys.\ B (Proc.\ Suppl.)\  {76} (1999) 245
[hep-ph/9810512].

\bibitem{RML}
T.~van Ritbergen, J.A.~Vermaseren and S.A.~Larin,
Phys.\ Lett.\ B {400} (1997) 379
[hep-ph/9701390];
%
K.G.~Chetyrkin,
Phys.\ Lett.\ B {404} (1997) 161
[hep-ph/9703278];
%
J.A.~Vermaseren, S.A.~Larin and T.~van Ritbergen,
Phys.\ Lett.\ B {405} (1997) 327
[hep-ph/9703284].

\bibitem{GLTC}
K.G.~Chetyrkin, A.L.~Kataev and F.~V.~Tkachov,
Phys.\ Lett.\ B {99} (1981) 147;
{\em ibid.}\ {101} (1981) 457 (E);
%
K.G.~Chetyrkin, S.G.~Gorishnii, S.A.~Larin and F.V.~Tkachov,
Phys.\ Lett.\ B {132} (1983) 351;
%
H.~Kleinert, J.~Neu, V.~Schulte-Frohlinde, K.G.~Chetyrkin and S.A.~Larin,
Phys.\ Lett.\ B {272} (1991) 39;
{\em ibid.}\ {319} (1991) 545 (E)
[hep-th/9503230];
%
B.~Kastening,
Phys.\ Rev.\ D {57} (1998) 3567
[hep-ph/9710346];
%
S.A.~Larin, M.~Monnigmann, M.~Strosser and V.~Dohm,
Phys.\ Rev.\ B 58 (1998) 3394
[cond-mat/9805028].

\bibitem{bn}
E.~Braaten and A.~Nieto,
Phys.\ Rev.\ Lett.\  {76} (1996) 1417
[hep-ph/9508406].

\bibitem{adjoint}
K.~Kajantie, M.~Laine, K.~Rummukainen and M.~Shaposhnikov,
Nucl.\ Phys.\ B {503} (1997) 357
[hep-ph/9704416].

\bibitem{az}
P.~Arnold and C.~Zhai,
Phys.\ Rev.\  {D 50} (1994) 7603
[hep-ph/9408276];
%
{\it ibid.}\  {51} (1995) 1906
[hep-ph/9410360];
%
C.~Zhai and B.~Kastening,
Phys.\ Rev.\  {D 52} (1995) 7232 [hep-ph/9507380];
%
E. Braaten and A. Nieto,
Phys.\ Rev.\ D 53 (1996) 3421 [hep-ph/9510408].

\bibitem{IR}
A.D.~Linde,
Phys.\ Lett.\ {B 96} (1980) 289.

\bibitem{dr}
P. Ginsparg, 
Nucl.\ Phys.\ B 170 (1980) 388;
%
T. Appelquist and R.D. Pisarski,
Phys.\ Rev.\ D 23 (1981) 2305.

\bibitem{a0cond}
K.~Kajantie, M.~Laine, K.~Rummukainen and Y.~Schr\"oder,
Phys.\ Rev.\ Lett.\  {86} (2001) 10
[hep-ph/0007109].

\bibitem{framework}
K.~Farakos, K.~Kajantie, K.~Rummukainen and M.~Shaposhnikov,
Nucl.\ Phys.\ {B 442} (1995) 317 [hep-lat/9412091].

\bibitem{contlatt}
M.~Laine,
Nucl.\ Phys.\ B {451} (1995) 484
[hep-lat/9504001];
%
M.~Laine and A.~Rajantie,
Nucl.\ Phys.\  {B 513} (1998) 471
[hep-lat/9705003].

\bibitem{moore_a}
G.D.~Moore,
Nucl.\ Phys.\  {B 493} (1997) 439 [hep-lat/9610013];
%
{\it ibid.}\  {523} (1998) 569
[hep-lat/9709053].


\bibitem{HS}
R.~Harlander and M.~Steinhauser,
Prog.\ Part.\ Nucl.\ Phys.\  {43} (1999) 167
[hep-ph/9812357].

\bibitem{FA}
J.~K\"ublbeck, M.~B\"ohm and A.~Denner,
Comput.\ Phys.\ Commun.\  {60} (1990) 165;
%
T.~Hahn,
hep-ph/0012260;
%
{\tt http://www.feynarts.de/}.

\bibitem{QG}
P.~Nogueira,
J.\ Comput.\ Phys.\  {105} (1993) 279;
%
{\tt ftp://gtae2.ist.utl.pt/pub/qgraf/}.

\bibitem{Klmany}
M.~Bachmann, H.~Kleinert and A.~Pelster,
Phys.\ Rev.\ D {61} (2000) 085017
[hep-th/9907044];
%
H.~Kleinert, A.~Pelster, B.~Kastening and M.~Bachmann,
Phys.\ Rev.\  {E 62} (2000) 1537
[hep-th/9907168];
%
B.~Kastening,
Phys.\ Rev.\  {E 61} (2000) 3501
[hep-th/9908172];
%
H.~Kleinert, A.~Pelster and B.~Van den Bossche,
hep-th/0107017;
%
A.~Pelster, H.~Kleinert and M.~Bachmann,
hep-th/0109014.

\bibitem{ma}
M.~Achhammer, 
{\em The QCD Partition Function at High Temperatures}, 
PhD thesis, University of Regensburg, July 2000 
(Logos-Verlag, Berlin, 2001).

\bibitem{inpre}
K.~Kajantie, M.~Laine, K.~Rummukainen and Y.~Schr\"oder,
in preparation. 
%

\bibitem{Cvi}
P.~Cvitanovi\'c, 
{\em Field Theory}, Nordita Lecture Notes 
(Nordita, Copenhagen, 1983).
%
{\em See also}
%
P.~Cvitanovi\'c, B.~Lautrup and R.B.~Pearson,
Phys.\ Rev.\ D {18} (1978) 1939.


\bibitem{LW}
J.M.~Luttinger and J.C.~Ward, 
Phys.\ Rev.\ {118} (1960) 1417;
%
G. Baym, 
Phys.\ Rev.\ {127} (1962) 1391;
%
C.~De~Dominicis and P.C.~Martin, 
J.\ Math.\ Phys.\ {5} (1964) 31.


\bibitem{CJT}
J.M.~Cornwall, R.~Jackiw and E.~Tomboulis,
Phys.\ Rev.\ D {10} (1974) 2428.

\bibitem{RS}
J.~Reinbach and H.~Schulz,
Phys.\ Lett.\ B {404} (1997) 291
[hep-ph/9703414].

\bibitem{effpot}
R.~Jackiw,
Phys.\ Rev.\ D {9} (1974) 1686;
%
R.~Fukuda and E.~Kyriakopoulos,
Nucl.\ Phys.\ B {85} (1975) 354.

\bibitem{CW}
S.~Coleman and E.~Weinberg,
Phys.\ Rev.\ D {7} (1973) 1888.

\bibitem{jamv}
J.A.M.~Vermaseren,
math-ph/0010025;
{\tt http://www.nikhef.nl/\~{}form/}.

\end{thebibliography}
\end{document}